\documentclass[12pt]{revtex4-1}
\usepackage{siunitx}
\usepackage{amsmath}  
\usepackage{amsfonts} 
\usepackage{graphicx}
\usepackage{array}
\usepackage{color}
\begin{document}

\title{A simple method for estimating the fractal dimension from digital images: The compression dimension}

\author{Pedro Chamorro-Posada}

\email{pedcha@tel.uva.es}
\affiliation{Departmento de Teor\'{\i}a de la Se\~nal y Comunicaciones e Ingenier\'{\i}a Telem\'atica, Universidad de Valladolid, ETSI Telecomunicaci\'on, Paseo Bel\'en 15, Campus Miguel Delibes, 47011 Valladolid, Spain} 
\date{\today}

\begin{abstract}
The fractal structure of real world objects is often analyzed using digital images.  In this context, the compression fractal dimension is put forward.  It provides a simple method for the direct estimation of the dimension of fractals stored as digital image files.  The computational scheme can be implemented using readily available free software.  Its simplicity also makes it very interesting for introductory elaborations of basic concepts of fractal geometry, complexity, and information theory.  A test of the computational scheme using limited-quality images of well-defined fractal sets obtained from the Internet and free software has been performed.  Also, a systematic evaluation of the proposed method using computer generated images of the Weierstrass cosine function shows an accuracy comparable to those of the methods most commonly used to estimate the dimension of fractal data sequences applied to the same test problem. 
\end{abstract}


\maketitle

\section{Introduction} 

One of the  most common descriptions of a given case study in Science and Engineering is through a graphical representation in an image.  Fractal analysis of digital images is of great value, for instance, in Medicine \cite{ahammer2001,ahammer2011,losa,waliszewski,cross} or Botanics \cite{castrejon2002,soil} and for the characterization of many other physical processes \cite{castrejon2003,berke}.

The algorithms normally used for calculating the fractal dimension of images \cite{ahammer,berke,spodarev} are rather involved and the absence of simple to use and freely distributed software tools can limit the widespread use of fractal methods in digital image analysis.  Furthermore, they are typically based on the processing of the image data that should be extracted from the picture file when this is the available source format.  In this work, a simple approach is proposed for estimating the information fractal dimension. The algorithm is oriented to its direct application to image files working with ordinary software tools and it can be used with no further prepossessing, no matter how the image is captured or generated.

The simplicity of the proposed computational scheme and its direct relation to basic concepts in information theory and complexity theory makes it suitable for computer lab experiments in fractal analysis \cite{hughes}. The potential of introductory fractal analysis even in high-school education was already highlighted in \cite{hurd}.

One major difference between the fractals found in empirical sciences and their mathematical counterparts is the existence of a finite limit to the scaling property in any real-world fractal. For fractal images, there are stringent restraints arising from the constrained image resolution \cite{ahammer2003} and the effect of noise \cite{reiss}.  To test the scheme proposed in this work, images of well-known fractal objects available in the Internet have been used without paying special attention to their resolution level.  Therefore, the results show the potential of the method for estimating the dimension of fractal images far from ideal conditions.  In a second systematic evaluation using computer synthesized fractals, the impact of the image resolution and other details of the implementation on the accuracy of the estimations is assessed.  The results of this study show that the performance of the proposed algorithm compares favorably, in terms of accuracy, with methods normally used for estimating the dimension of fractal sequences.

\section{The compression dimension}

\subsection{Information fractal dimension}

Hausdorff dimension provides a rigorous mathematical definition of dimension \cite{mandelbrot}.  In an intuitive way, this concept can be introduced through the exponent describing the variation of the size of an object with the scale used to measure it \cite{mandelbrot,theiler}, 
\begin{equation}
\text{size}\sim\text{scale}^\text{dimension}.
\end{equation}
For a segment, both its size and scale are given by its length and the dimension is one. A circle is an example of a two-dimensional object since its size (area) scales with its diameter as $\text{size}=\pi\times\text{scale}^2$.  For a sphere the size (volume) is related with the scale (diameter) as $\text{size}=\pi/6\times\text{scale}^3$ and its dimension is three. A fractal object in the plane, like a coastline, will have dimension larger than one (and smaller than two) as a consequence of the space-filling properties of the graph and its infinite length. 

Calculating fractal dimensions is the primary objective in the study of fractals and can be a fairly complex task.  One possibility for calculating fractal dimensions is the box-counting approach.  At each resolution $r$, one defines a grid covering the object that is being analyzed (squares for the plane and cubes in space) and then counts the number $n(r)$ of nonempty grid boxes.  The box-counting dimension is then defined as
\begin{equation}
D_B=\lim_{r\to 0}\frac{-\log n(r)}{\log r}=\lim_{s\to \infty}\frac{\log n(s)}{\log s},
\end{equation}
or $n(r)\sim \left(1/r\right)^{D_B}$, i.e. $n(s)\sim s ^{D_B}$, where the scale $s$ is the inverse of the resolution $s=1/r$.

An alternative approach is given by the information dimension  \cite{theiler}.  One determines how many bits of information $H(r)$ are needed to specify a point in the object with a accuracy set by $r$.  The information dimension is then given by
\begin{equation}
D_I=\lim_{r\to 0}\frac{-H(r)}{\log r}=\lim_{s\to \infty}\frac{H(s)}{\log s}. \label{eq:DI}
\end{equation}
$H$ is the Shannon Entropy \cite{cover} of the fractal.  If we partition the fractal in boxes of size $r$ we need
\begin{equation}
H=-\sum_{i} P_i \log_2 P_i \label{entropia}
\end{equation}
bits of information to specify one box or, equivalently, to specify the position of a point in the fractal to an accuracy $r$.  $P_i$ in (\ref{entropia}) is the probability measure (the size) of  box $i$.  

Different indirect estimates for the entropy have been used to analyze data sequences in complex dynamical systems, such as electroencephalograms \cite{kannthal}.  Our approach, instead, focuses on the direct estimation of the information dimension of geometrical objects based on data compression.

\subsection{Data compression}

Data compression aims to produce an encoding that gives the shortest possible description of the information content of the data.  Shannon entropy is the fundamental lower bound for compressing information\cite{cover}. For the commonly employed compression schemes, like Lempel-Ziv (LZ) algorithm, it can be proved  \cite{cover} that the compressed file size equals the entropy of the data asymptotically in the number of symbols.  From a practical point of view, such assumption is reasonable for regular image file sizes.  A second compression of an efficiently compressed file should yield a negligible size reduction ratio, since the file size is already very close to the entropy limit.  This can be used as a check for the performance of a file compression software.  Also, the entropy limit can be approached in a two-step scheme if the compression efficiency is poor for large files.  

We will use freely available and very efficient data compression software to obtain approximate values of Shannon entropy in our calculations of the fractal dimension.  Data compression software is also routinely used, for instance, for estimating the Kolmogorov complexity distance \cite{kaitchenko}. 

Note that we have to use lossless data compression algorithms that permit us to fully recover the uncompressed data, and we have to be careful to avoid lossy compression algorithms used, for instance, in JPEG image file files that achieve high compression rates at the cost of loss of information.

\subsection{Image files}

 There are two types of graphic formats for the computer representation of images.  In \emph{vector} graphic formats the different elements that constitute the image are mathematically specified as geometrical primitives (such as lines, circles, etc.).  Therefore, the image file contains indications for reconstructing the image at any required level of detail.  Scaling an image stored in a vector graphics file is a reversible operation and it does not affect the amount of information required to describe the image.  In \emph{raster} (or \emph{bitmap}) graphics formats, on the other hand, an image is stored as a matrix of pixels.   As we decrease the resolution of the raster image we disregard  image pixels, there is a loss of information with the result that the image cannot be recovered to the previous level of detail from the reduced scale, and the amount of information required for describing the image decreases in accordance to the reduction in the complexity.  
  
	There are many possible choices for the bitmap graphics format with different compression options \cite{Salomon}.  For instance, the Graphics Interchange Format (GIF) is widely used in the Internet.  The GIF format uses LZW data compression, whereas the also commonly employed Portable Network Graphics (PNG) format is based on the DEFLATE compression algorithm.  Both compression methods are lossless and belong to the class of dictionary compression methods of the LZ method that share the entropy property.  This means that an efficient implementation would give a compressed file size asymptotically approaching the entropy of the data \cite{Salomon}.  In the Tagged Image File Format (TIFF) one can choose among lossy JPEG image compression, several types of lossless compression or no compression at all.  All the compression methods employed in this study: DEFLATE  (in PNG image format, in ZIP compressed TIFF format and in the {\tt gzip} software for external compression of files) and LZW (in TIFF format) belong to the class of lossless dictionary compression methods \cite{Salomon}.

\subsection{The compression dimension}

Similarly to the definition of information dimension of a fractal object, we now consider the scaling effect on compressed image files of its pictorial representation. If we use an image with $N_i=n_x\times n_y$ pixels we need $N_i$ symbols to store it, one per pixel.  After compression, the minimum file size $S$, expressed in bits, required to store this information is \cite{cover}
\begin{equation}
S=N_i h, \label{eq:size} 
\end{equation}
where $h$ (bits/symbol) is the entropy rate of the data file. $S$ is, by definition \cite{cover}, the entropy of the data file. 

We now define a magnifying factor of the image (or scale) $s$ such as the total number of pixels used to represent the image is
\begin{equation}
N_i(s)=n_x\times n_y=(sn_{0x})\times (sn_{0y})=N_0 s^2, \label{eq:scale}
\end{equation} 
with $N_0=n_{0x}\times n_{0y}$ an arbitrary reference value of the number of pixels.

The optimal compressed file size used for storing the image is, using \eqref{eq:size} and \eqref{eq:scale},
\begin{equation}
S(s)=N_0s^2h(s).\label{eq:S}
\end{equation}

In our former definitions \eqref{eq:DI}, \eqref{eq:size}, $H$  applies strictly to the fractal set and $S$ and $h$ to the image file.  Now, we develop the relationship that exists between the entropy of the fractal and that of its image representation.     At a scale $s$, $H(s)$ is the number of bits required to specify a point of the fractal.  Therefore, the total number of points required to specify the fractal at scale $s$, $N_f(s)$, according to \eqref{eq:DI}, is  
\begin{equation}
 N_f =2^{H(s)}=N_1s^D,
\end{equation}
where $N_1$ is an unknown integer, since $s$ has been arbitrarily referenced to $N_0$. 

We consider a black and white image  of the fractal where an image pixel is coded with the bit $1$ if it corresponds to a fractal point and bit $0$ otherwise. For a \emph{faithful representation}, we need that the number of available image pixels exceeds largely the number of pixels required for the description of the fractal at a given resolution level $N_f(s)<<N_i(s)$.  A lossless compression of this image can be obtained using the \emph{run-length encoding algorithm} \cite{Salomon}.  First, the image can be represented as a binary sequence obtained by the concatenation of the image rows.  Then, the information in the image can be encoded as the positions of the $1$ bits within that sequence.  The list of the stored position values of the black pixels can then be further compressed using a conventional Huffman encoding \cite{Salomon}.  In the conditions specified above, the entropy of the image file obeys the scaling asymptotics
\begin{equation}
S(s)\sim 2^{H(s)}\sim s^D. \label{eq:dim1} 
\end{equation}    

The above equation serves as a definition of the compression dimension of a fractal $D_C$ as
\begin{equation}
D_C\equiv \lim_{s\to \infty}\dfrac{\log S(s)}{\log s},
\end{equation}
and we expect that $D_C$ permits to estimate the value of $D$.

\section{Computational procedure}\label{algoritmo}

The computational procedure used in this work is now described.  Of course, this recipe can be conveniently adapted to any particular scenario.   We will use compressed image file representations of an object at different scales in order to estimate its fractal dimension. Any lossless type of data compression, either included in the coded bitmap image file or external to it can be used for this purpose.  For our first test experiment, we start with an uncompressed TIFF image at each scale and we compress it using {\tt gzip}.  We have checked that  using PNG graphics format without further compression produces very similar results.

\begin{figure}
\begin{tabular}{cc}
(a)&(b)\\
\includegraphics[width=8cm]{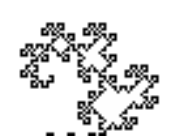}&\includegraphics[width=8cm]{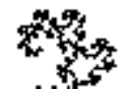}\\
(c)&(d)\\
\includegraphics[width=8cm]{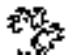}&\includegraphics[width=8cm]{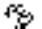}\\
\end{tabular}
\caption{A small portion of the boundary of the Dragon curve shown in Figure \ref{dragon} corresponding to the original image (a) and at $s=7$ (b) $s=4$ (c) and $s=2$ (d).}\label{escalas}.
\end{figure}

In the first test experiment, the computational procedure used for estimating the fractal dimension is as follows:

\begin{itemize}
\item STEP 0:  Generate an initial uncompressed TIFF version of the downloaded image.  Since the source image for which the fractal dimension is estimated can have any type of image format encoding, this step has the specific purpose of uniforming the estimation procedure.
\item STEP 1:  Generate nine versions of the fractal image as TIFF files with no compression at different scales $s=1,2,\dots9$.  This corresponds to reducing the image size to $10\%,20\%,\cdots,90\%$ of the original file size. 
\item STEP 2:  Compress all the tiff files.
\item STEP 3:  Measure the file sizes $S(s)$  and plot $\log(S)$ versus $\log(s)$ and determine the physical scaling range.
\item STEP 4:  Determine, using linear regression, the slope of the log-log plot.  This is the estimated value of the fractal dimension $D$ since
\[
S\sim s^D.
\] 
\end{itemize}

\begin{figure}
\begin{tabular}{cc}
(a)&(b)\\
\includegraphics[width=8cm]{trozoimagen.pdf}&\includegraphics[width=8cm]{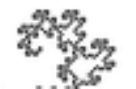}\\
(c)&(d)\\
\includegraphics[width=8cm]{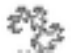}&\includegraphics[width=8cm]{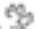}\\
\end{tabular}
\caption{A small portion of the boundary of the Dragon curve shown in Figure \ref{dragon} corresponding to the original image (a) and at $s=7$ (b) $s=4$ (c) and $s=2$ (d) for gray level representations.}\label{escalasgris}.
\end{figure}

The free software image processing suit \verb/imagemagick/  \cite{imagick} has been used for step $1$. For instance, the command
\begin{verbatim}
convert -resize 10% -monochrome -compress None Image.tiff image_s_1.tiff
\end{verbatim}
permits one to obtain the smallest $s=1$ representation of the original image in file \verb/Image.tiff/ in the file \verb/image_s_1.tiff/ by resizing the image to a $10\%$ of its original size keeping the image as a black and white (monochrome) image and using no compression.  For step $2$, the free compression software \emph{GNU zip} ({\tt gzip})\cite{gzip} has been used.

In the second experiment, different image formats and compression types have been studied.  For this reason, STEPs $1$ and $2$ have been merged in a single operation using {\tt ImageMagick} software.  Also, a second compression of the files has been performed using {\tt gzip}.  This has permitted to identify situations where the efficiency of the compression was poor and improve the accuracy of the results in these cases.  

Figure \ref{escalas} displays a small portion of the Dragon fractal curve \ref{figswiki} at the original and three different scaling levels.  We can see how changing the pixel size is, in some sense, related with the change of the box resolution $r$ in a box counting experiment, but with one notable difference:  when the scale is reduced, a given pixel (box) is determined to be filled or not by sampling the former image, which produces an additional loss of information.  The use of gray images in the scaling of the original image, as illustrated in figure \ref{escalasgris} for the same case, can solve this issue.   Now, each pixel is not only either black or white, but it can have any in a large number of intermediate gray values.  The particular gray value is related to the number of black and white pixels in the area of the original image that is collapsed to this particular pixel in the scale reduction process. Therefore, grayscale images can actually be advantageous for calculating the fractal dimension since the loss of information due to sampling in the rescaling process is avoided.  This difference between the amount of information given by BW or gray images is also related with one of the main limitations of the box-counting algorithm in practical applications that has led to the definition of a generalized box-counting dimension \cite{theiler}.  In this scheme, boxes are not simply occupied or not by the object, but the number of occupied points in a box are considered, much like in a grayscale image. 

The {\tt -resize} command in {\tt ImageMagick} has many options that affect how the downscaled images are calculated differently depending on the image format \cite{imagick}.  For this  reason, a simplified version provided by the command {\tt -scale} as a fixed pixel averaging procedure \cite{imagick}, has been used in the second test experiment for changing the scale of the images consistently among the various image formats.

\section{Results and discussion}

The proposed method has been applied to two different case studies.  First, various images of fractal sets downloaded from the Internet have been analyzed.  Then, a systematic survey based on the Weierstrass fractal has been used to draw more general conclusions regarding the properties of the method.

\subsection{Downloaded image files}

Even though the underlying objects in the first part of the study are precisely defined mathematical fractals, the image files we work with are \emph{real world} fractals and the level of detail in the original image permits only a finite depth in the scaling procedure.  For instance, the image analyzed in figure \ref{escalas}, at $s=1$ is completely blank.  Therefore, \verb/STEP 3/ includes the study of the scaling plot to determine the scaling range of interest.  This can typically be identified from a change of the slope in the graph.

\begin{figure}
\begin{tabular}{>{\centering\arraybackslash}m{8cm}>{\centering\arraybackslash}m{8cm}}
(a)&(b)\\
\includegraphics[width=8cm]{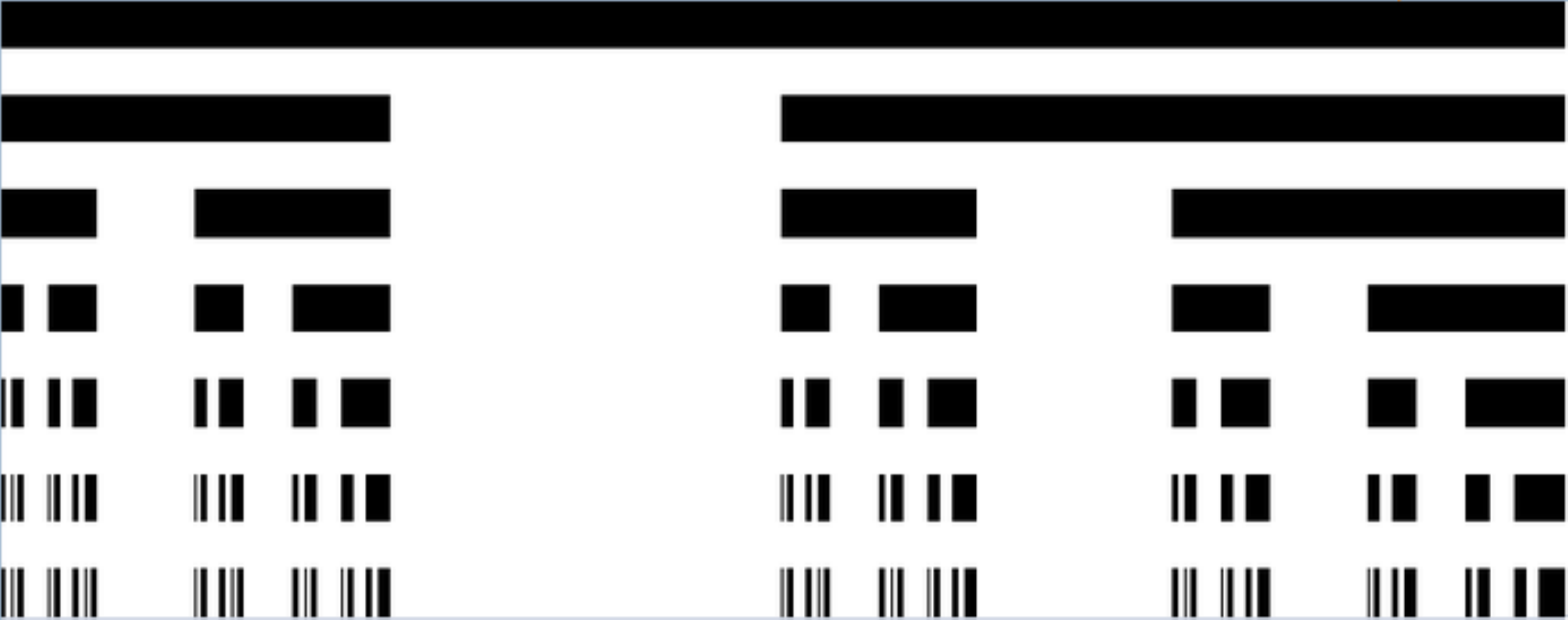}&\includegraphics[width=8cm]{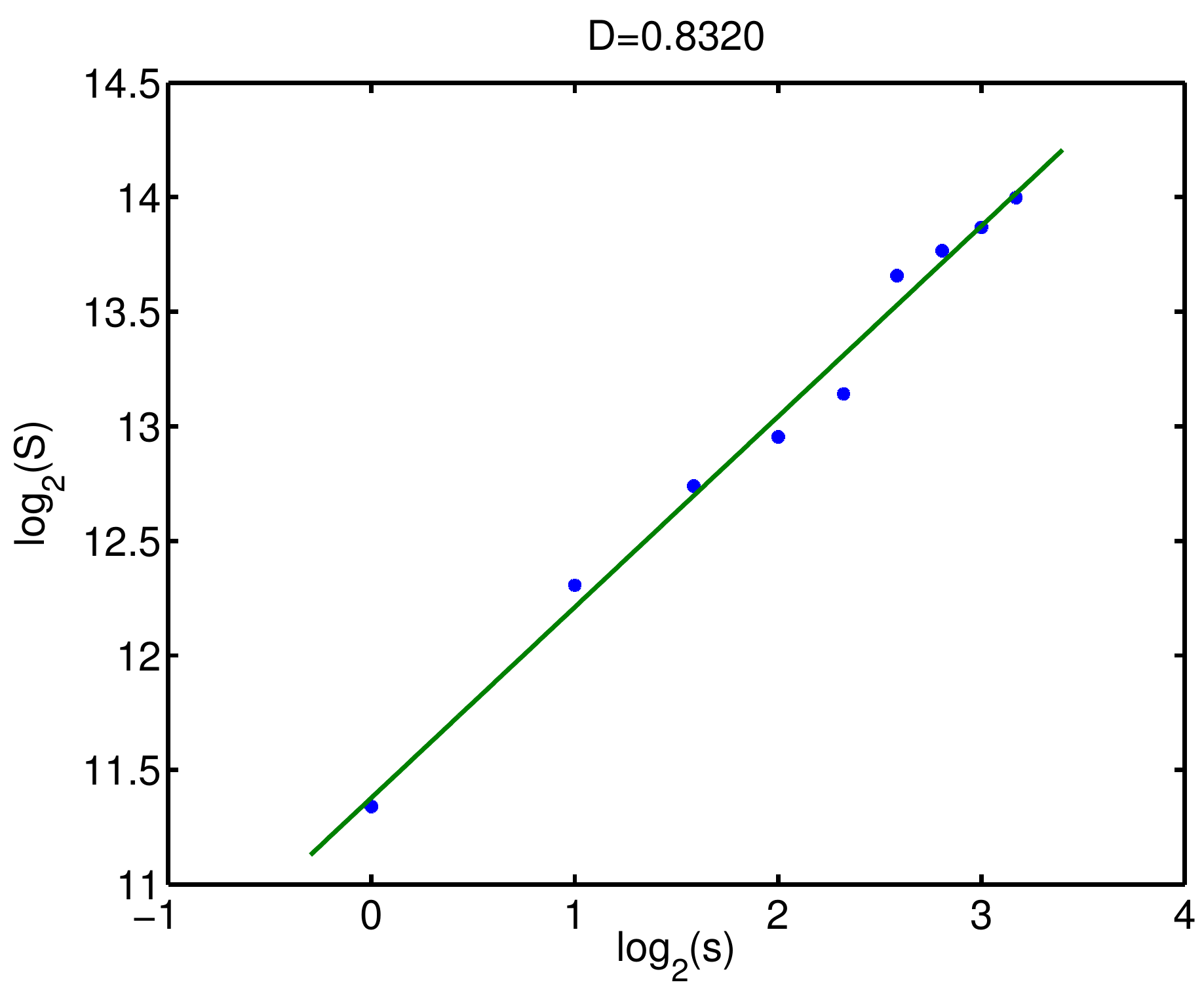}
\end{tabular}
\caption{(a) Asymmetric Cantor set and (b) its fractal dimension analysis.}\label{cantor}
\end{figure}
\begin{figure}
\begin{tabular}{>{\centering\arraybackslash}m{8cm}>{\centering\arraybackslash}m{8cm}}
(a)&(b)\\
\includegraphics[width=8cm]{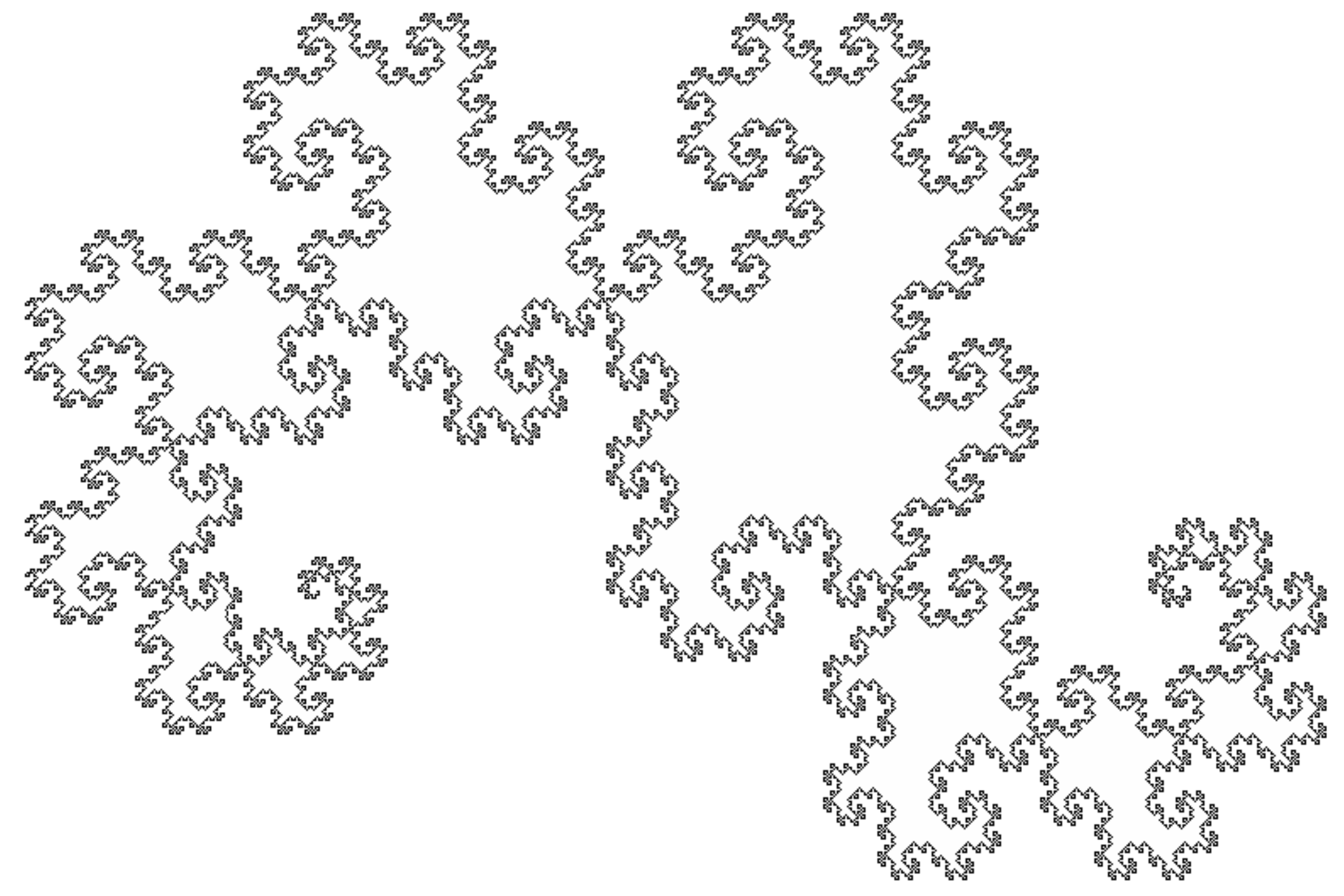}&\includegraphics[width=8cm]{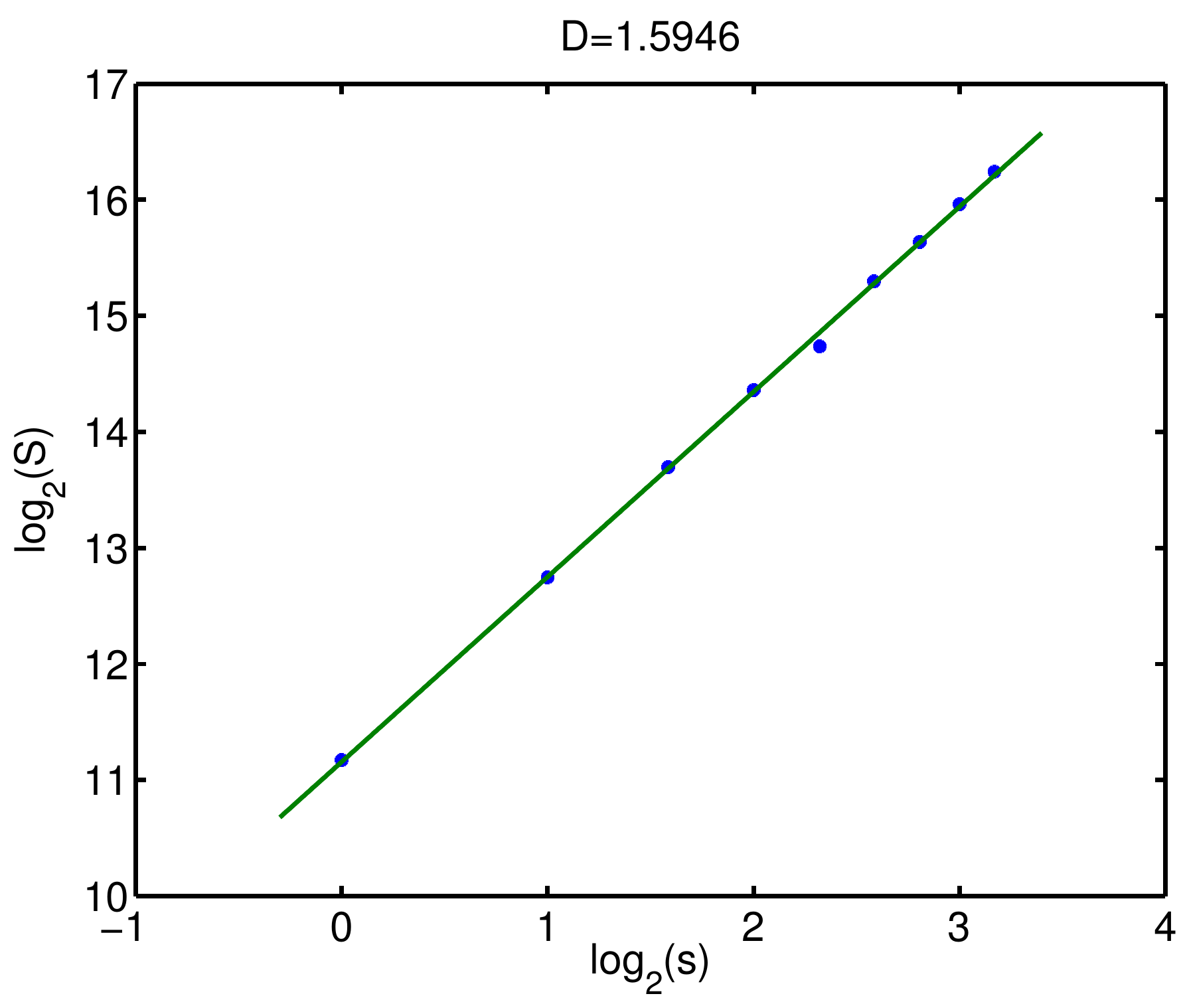}\\
\end{tabular}
\caption{(a) Boundary of the Dragon curve and (b) its fractal dimension analysis.}\label{dragon}
\end{figure}
\begin{figure}
\begin{tabular}{>{\centering\arraybackslash}m{8cm}>{\centering\arraybackslash}m{8cm}}
(a)&(b)\\
\includegraphics[width=8cm]{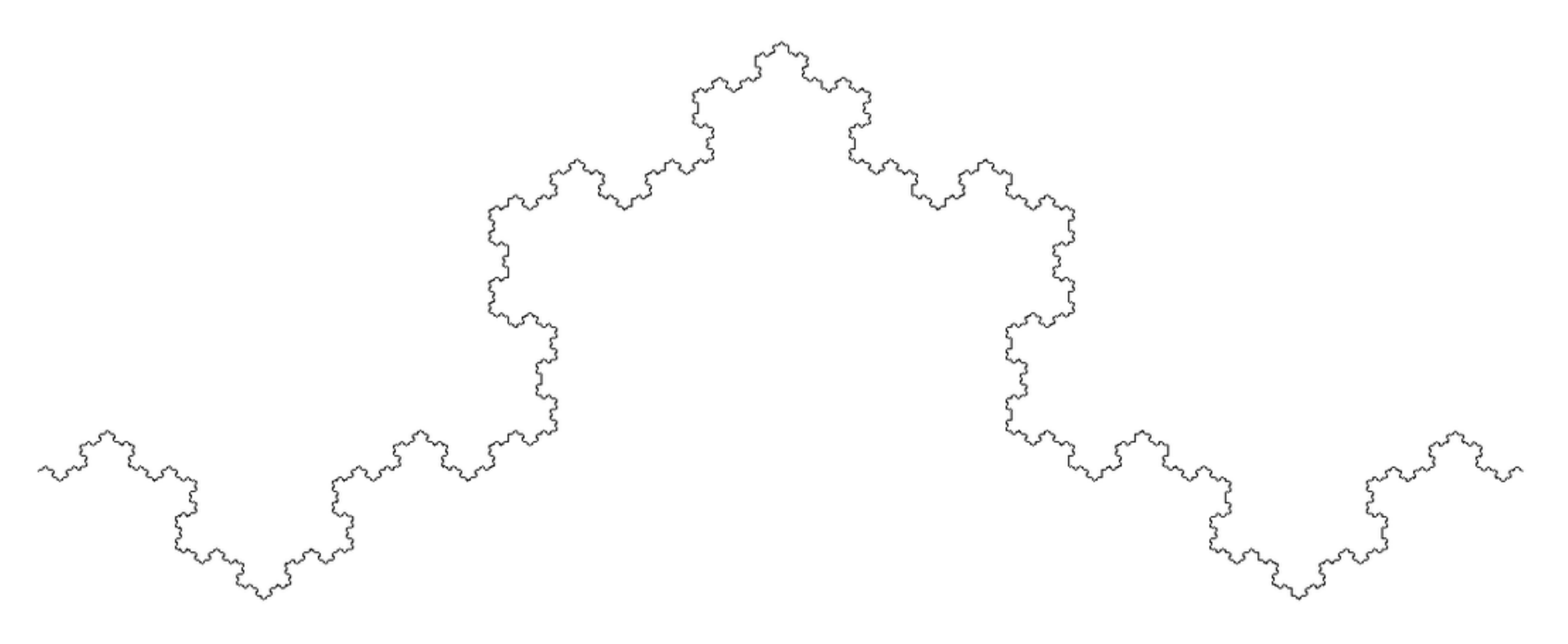}&\includegraphics[width=8cm]{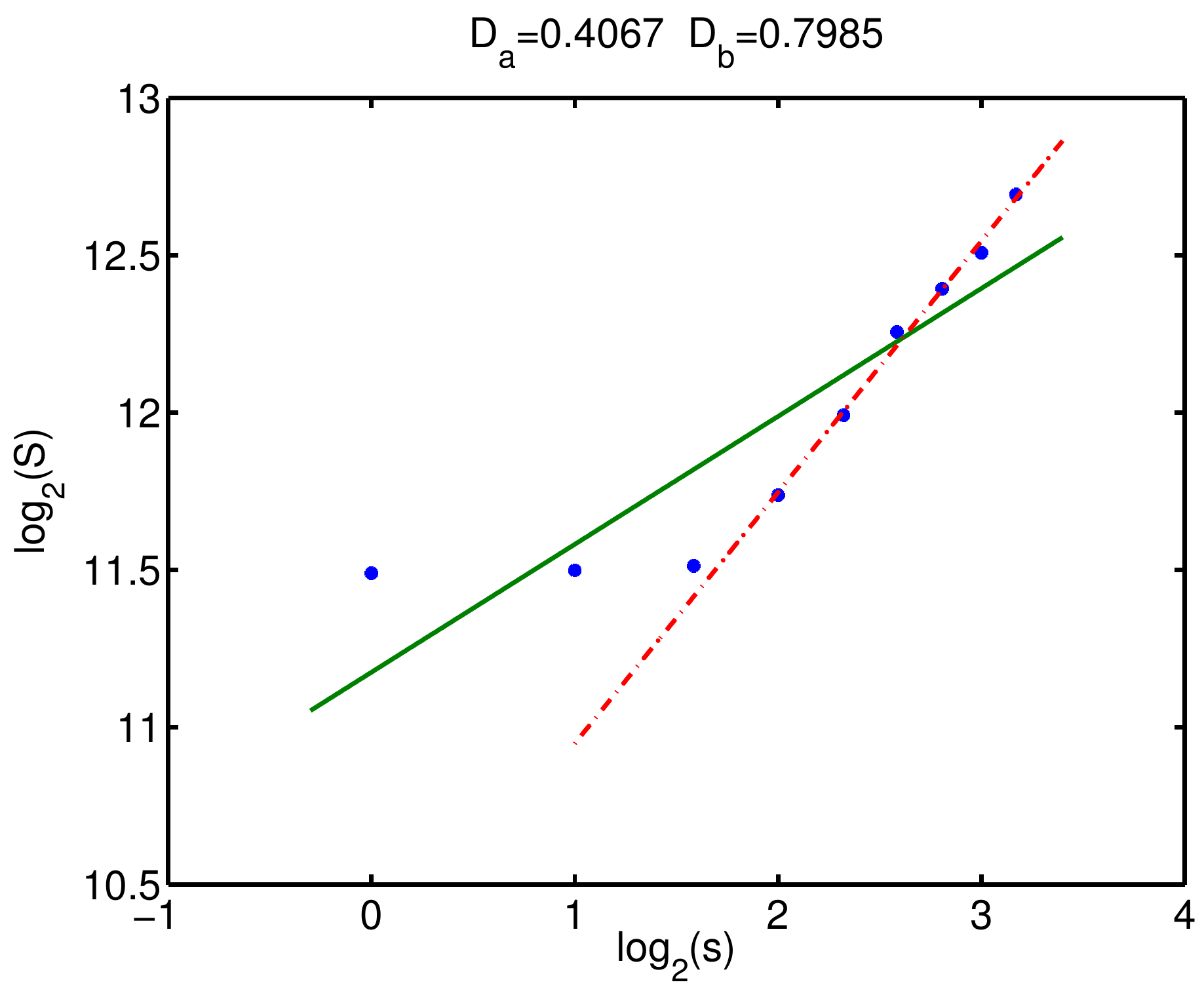}\\
\end{tabular}
\caption{(a) Fibonacci word fractal $60^o$ and (b) its fractal dimension analysis.}\label{fibo}
\end{figure}
\begin{figure}
\begin{tabular}{>{\centering\arraybackslash}m{8cm}>{\centering\arraybackslash}m{8cm}}
(a)&(b)\\
\includegraphics[width=8cm]{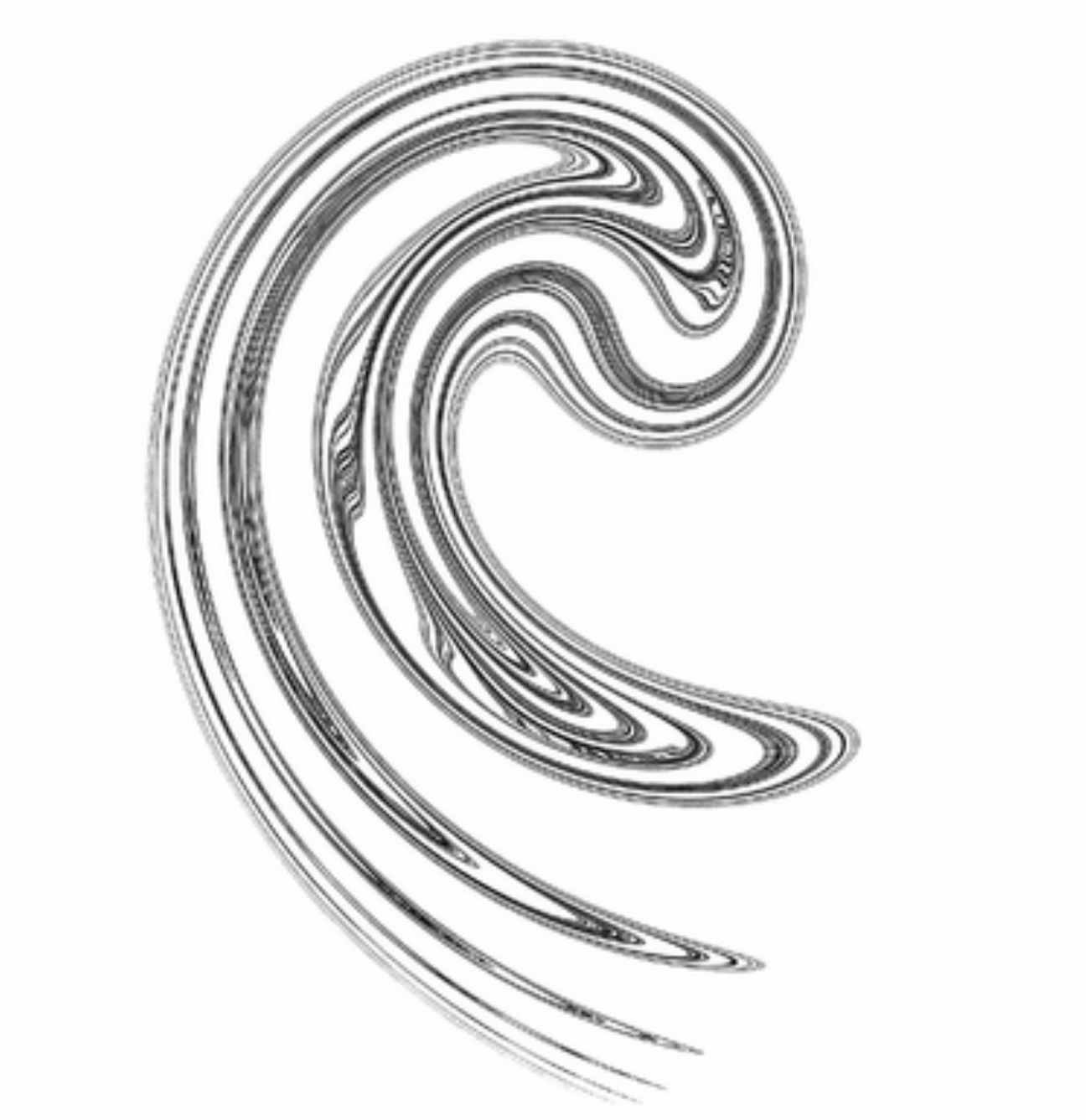}&\includegraphics[width=8cm]{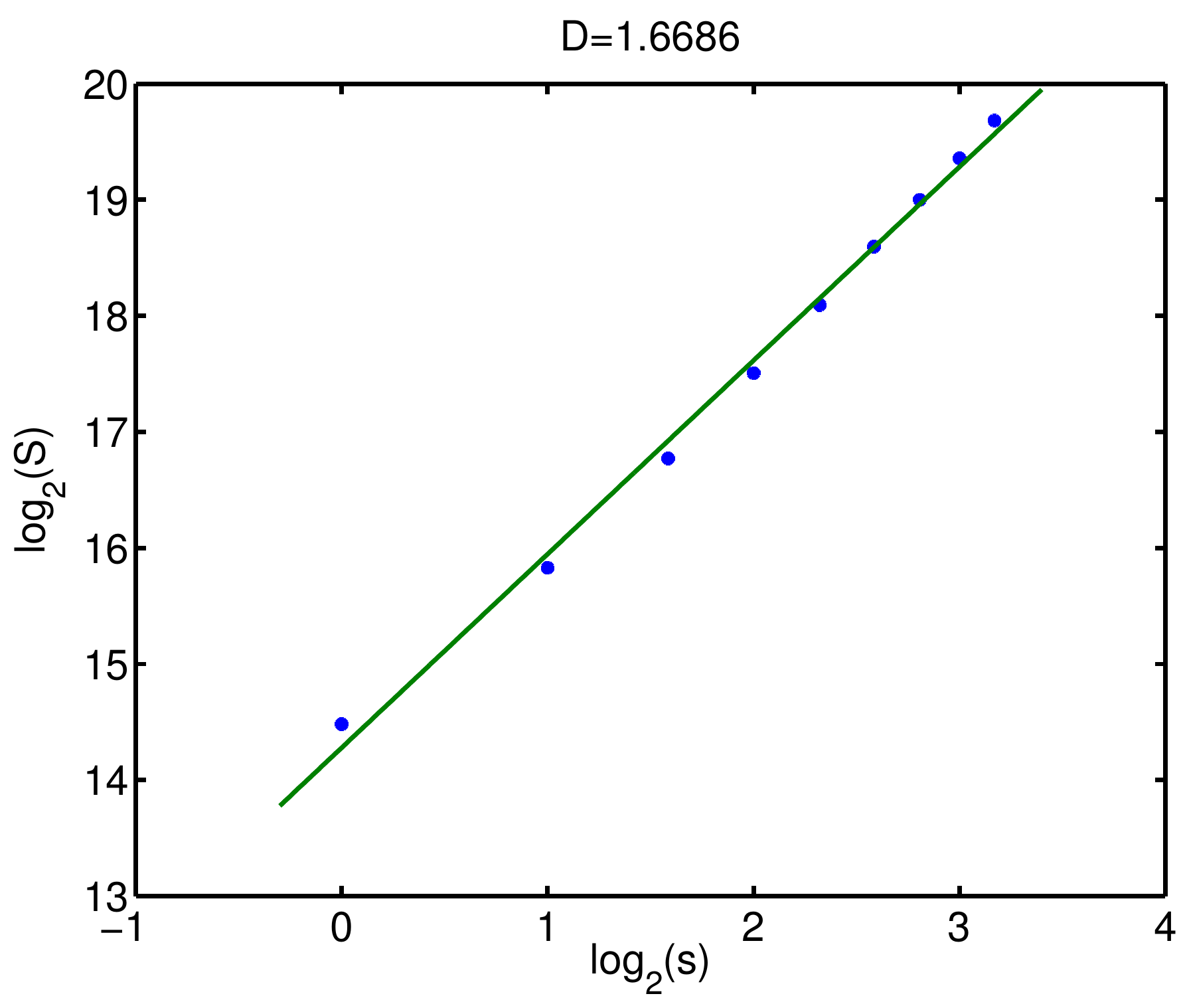}\\
\end{tabular}
\caption{(a) Ikeda map attractor and (b) its fractal dimension analysis.}\label{ikeda}
\end{figure}
\begin{figure}
\begin{tabular}{>{\centering\arraybackslash}m{8cm}>{\centering\arraybackslash}m{8cm}}
(a)&(b)\\
\includegraphics[width=8cm]{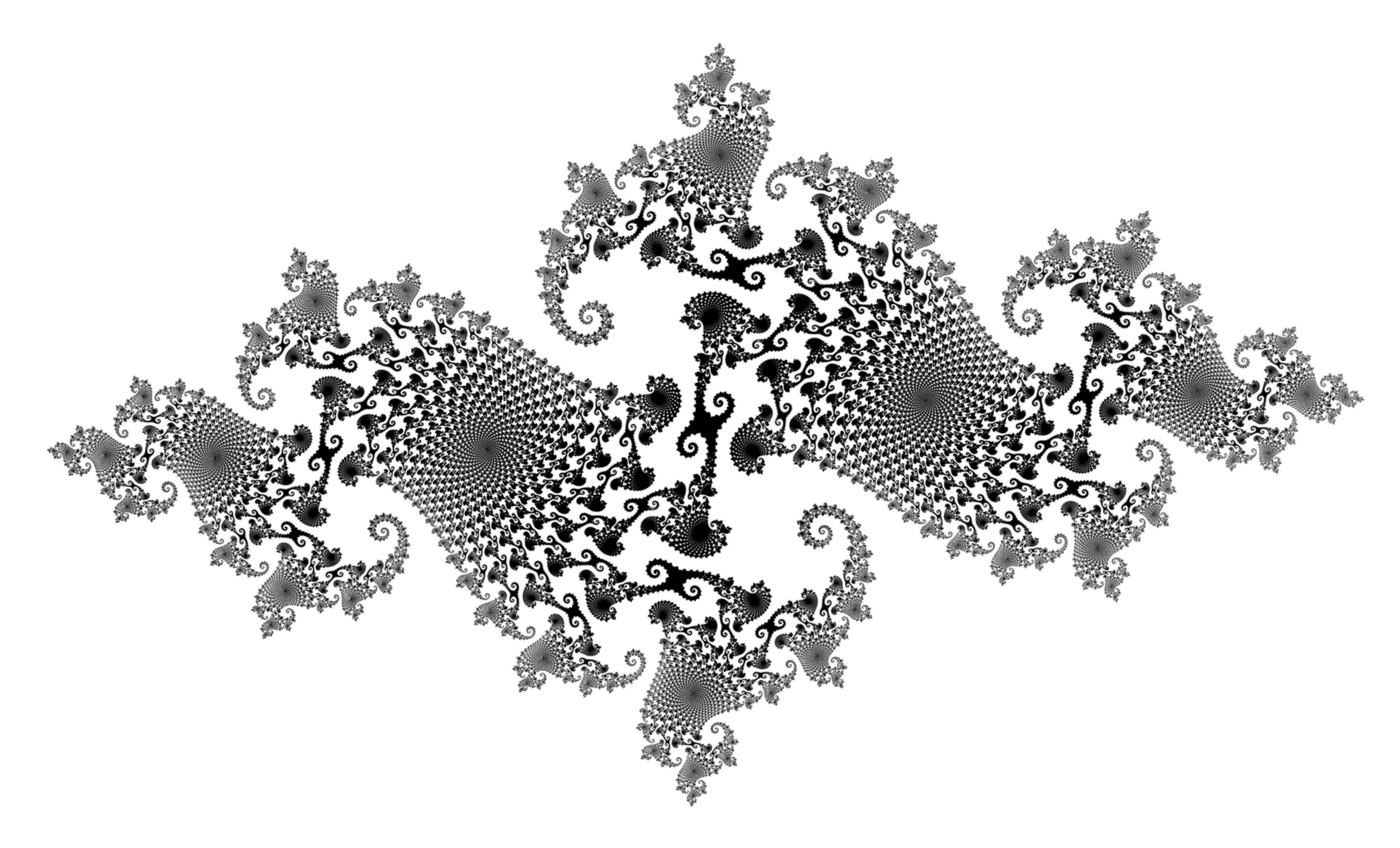}&\includegraphics[width=8cm]{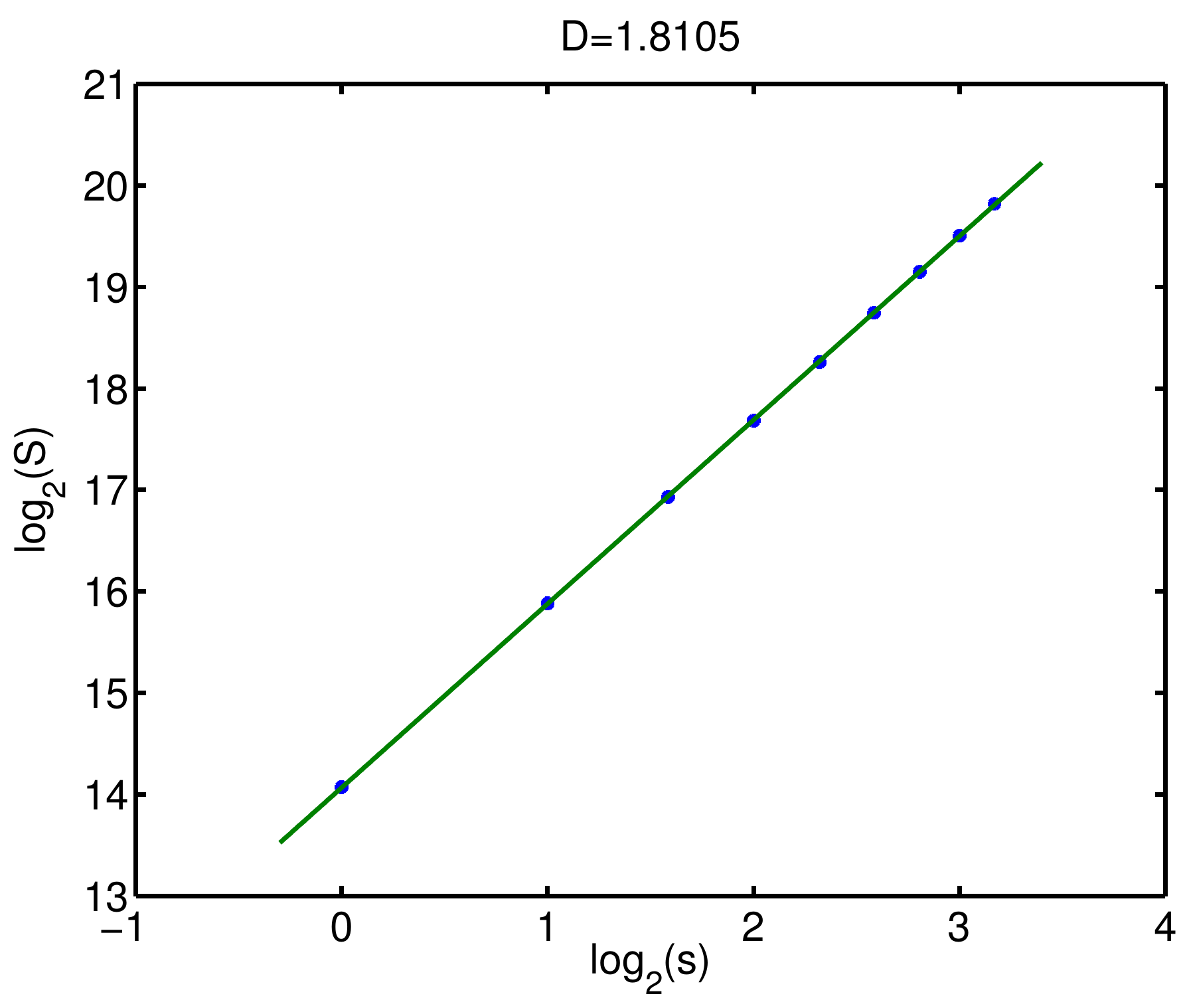}\\
\end{tabular}
\caption{(a) Julia set and (b) its fractal dimension analysis.}\label{julia}
\end{figure}
\begin{figure}
\begin{tabular}{>{\centering\arraybackslash}m{8cm}>{\centering\arraybackslash}m{8cm}}
(a)&(b)\\
\includegraphics[width=8cm]{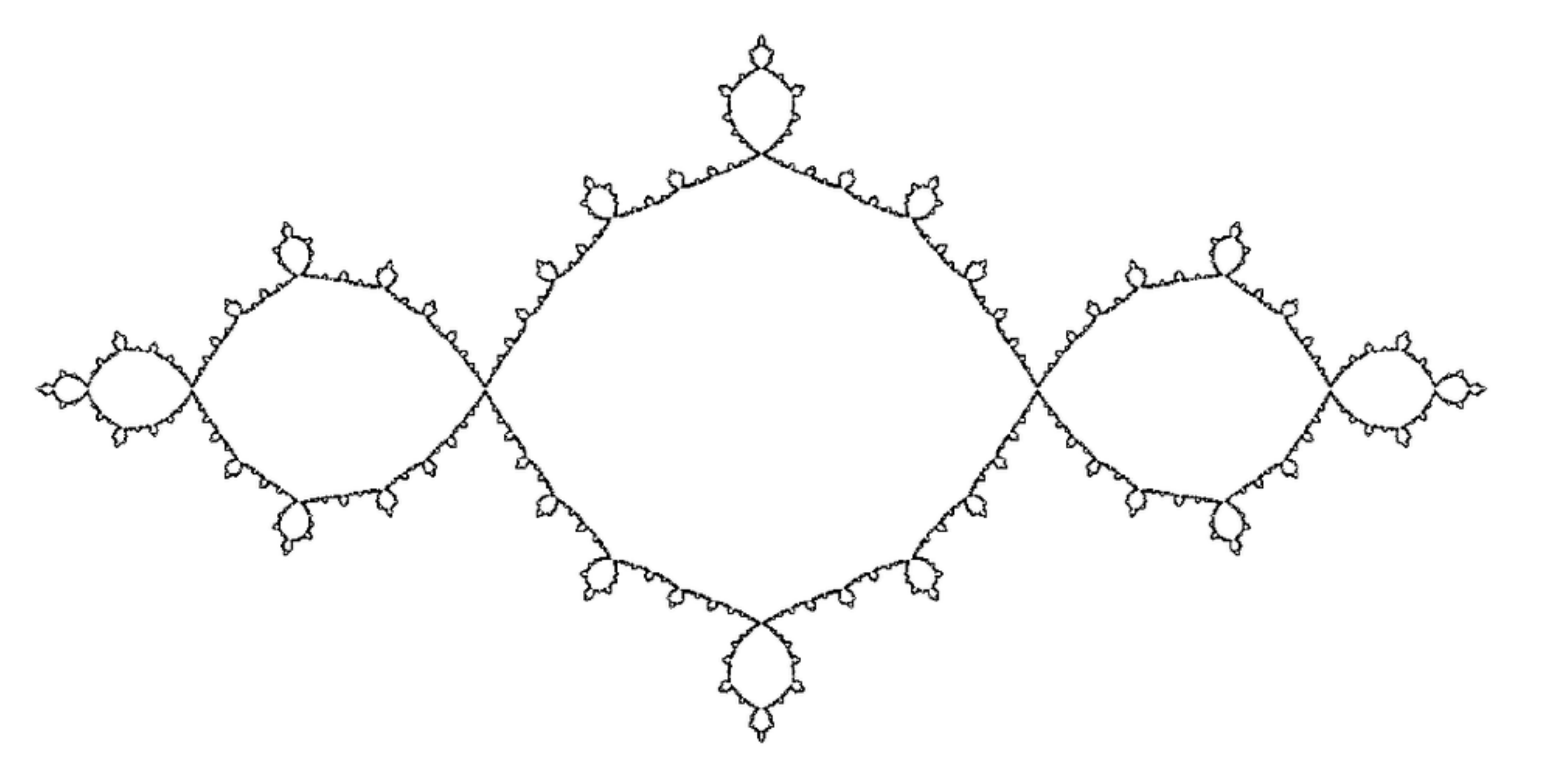}&\includegraphics[width=8cm]{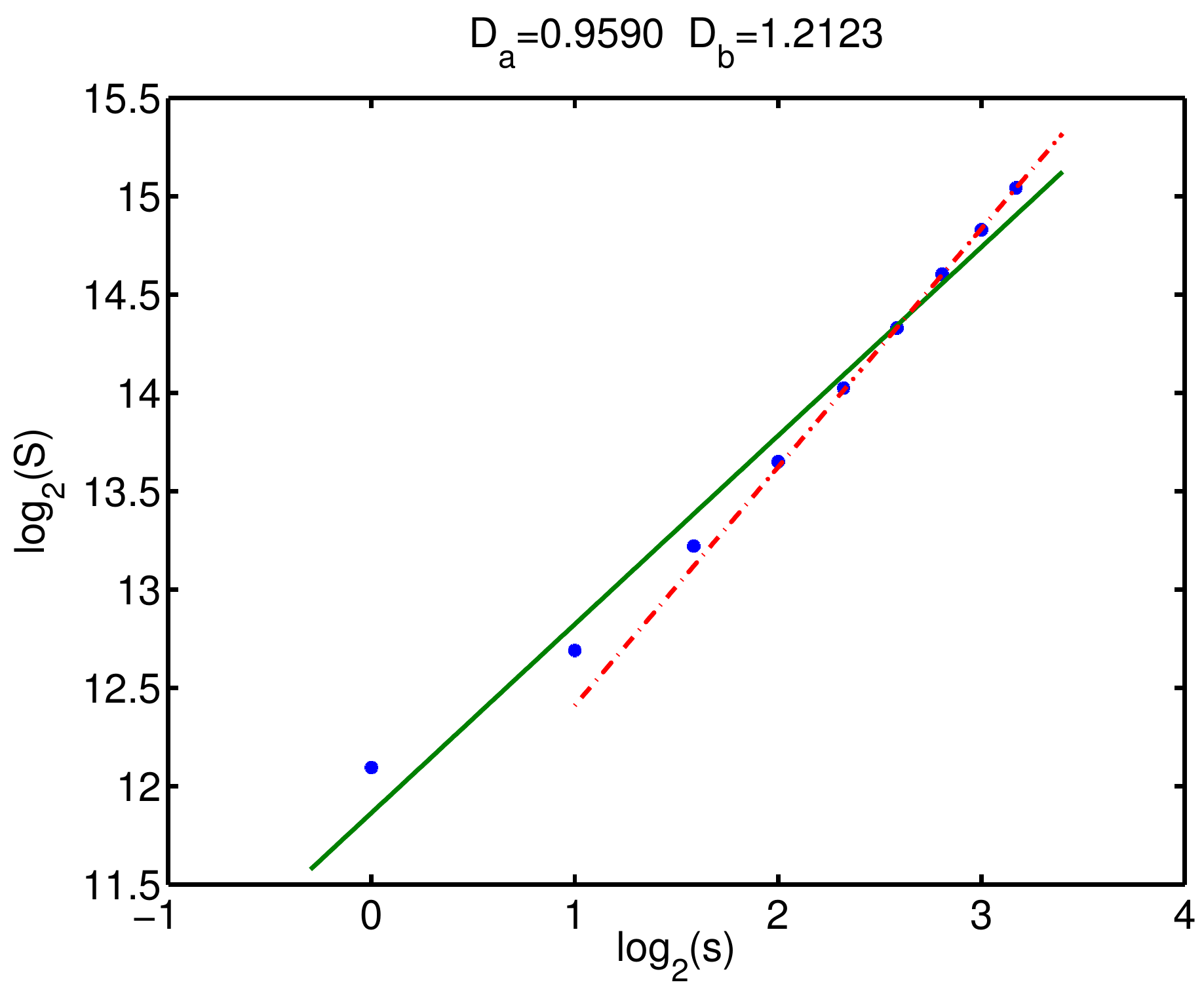}\\
\end{tabular}
\caption{(a) Julia set $z^2-1$ and (b) its fractal dimension analysis.}\label{juliaz}

\end{figure}
\begin{figure}
\begin{tabular}{>{\centering\arraybackslash}m{8cm}>{\centering\arraybackslash}m{8cm}}
(a)&(b)\\
\includegraphics[width=8cm]{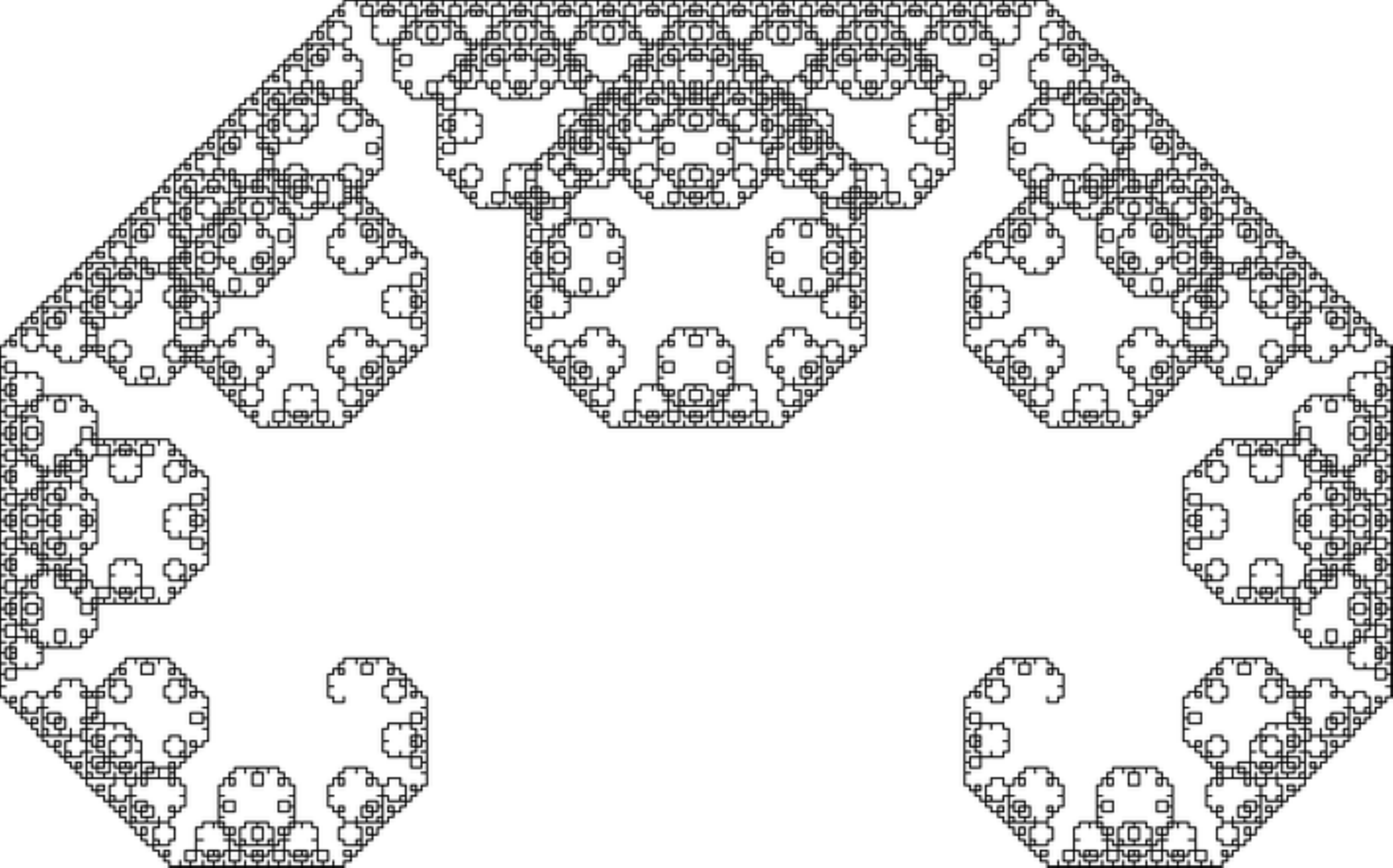}&\includegraphics[width=8cm]{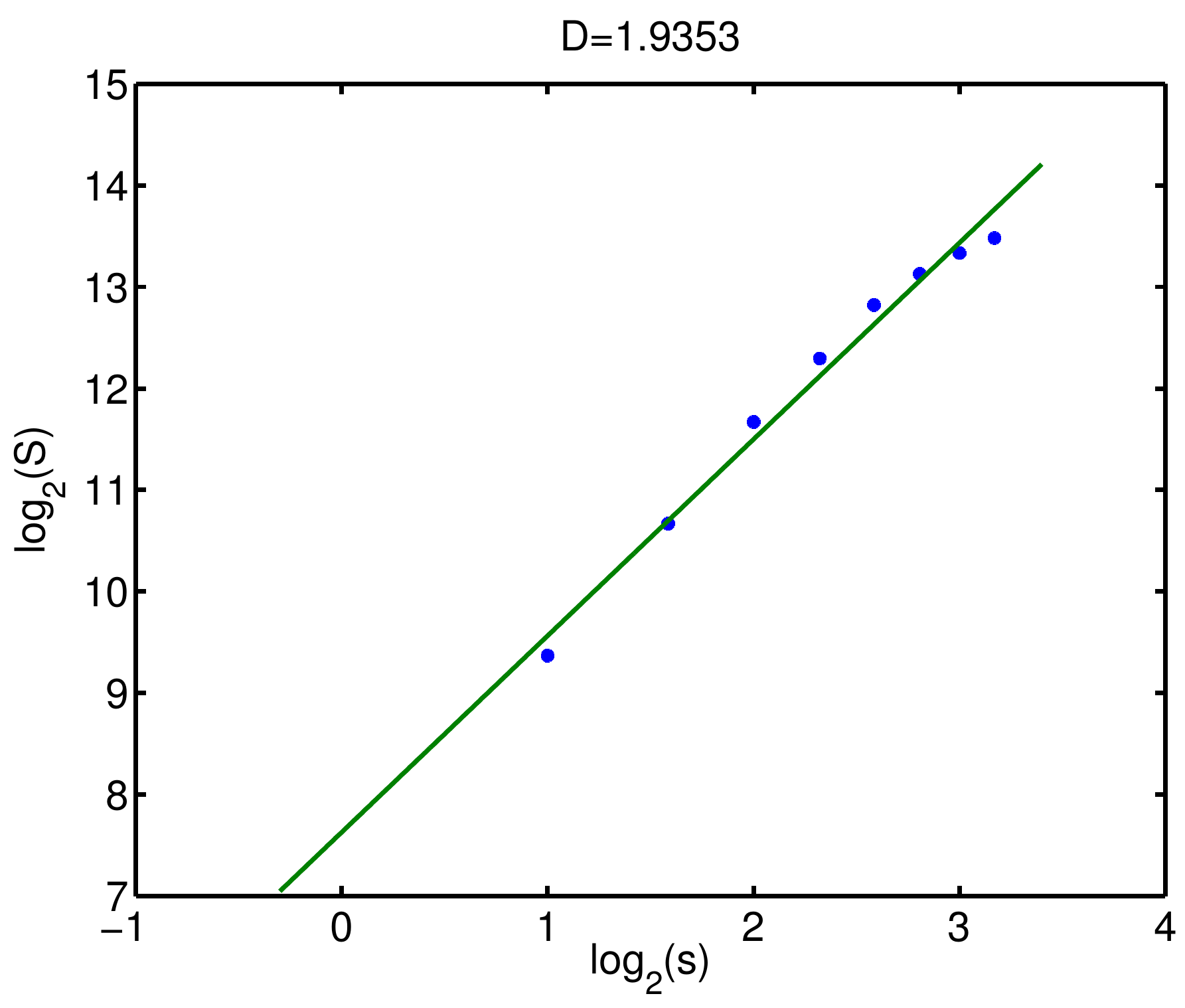}\\
\end{tabular}
\caption{(a) Boundary of the L\'evy C curve and (b) its fractal dimension analysis.}\label{levy}

\end{figure}
\begin{figure}
\begin{tabular}{>{\centering\arraybackslash}m{8cm}>{\centering\arraybackslash}m{8cm}}
(a)&(b)\\
\includegraphics[width=8cm]{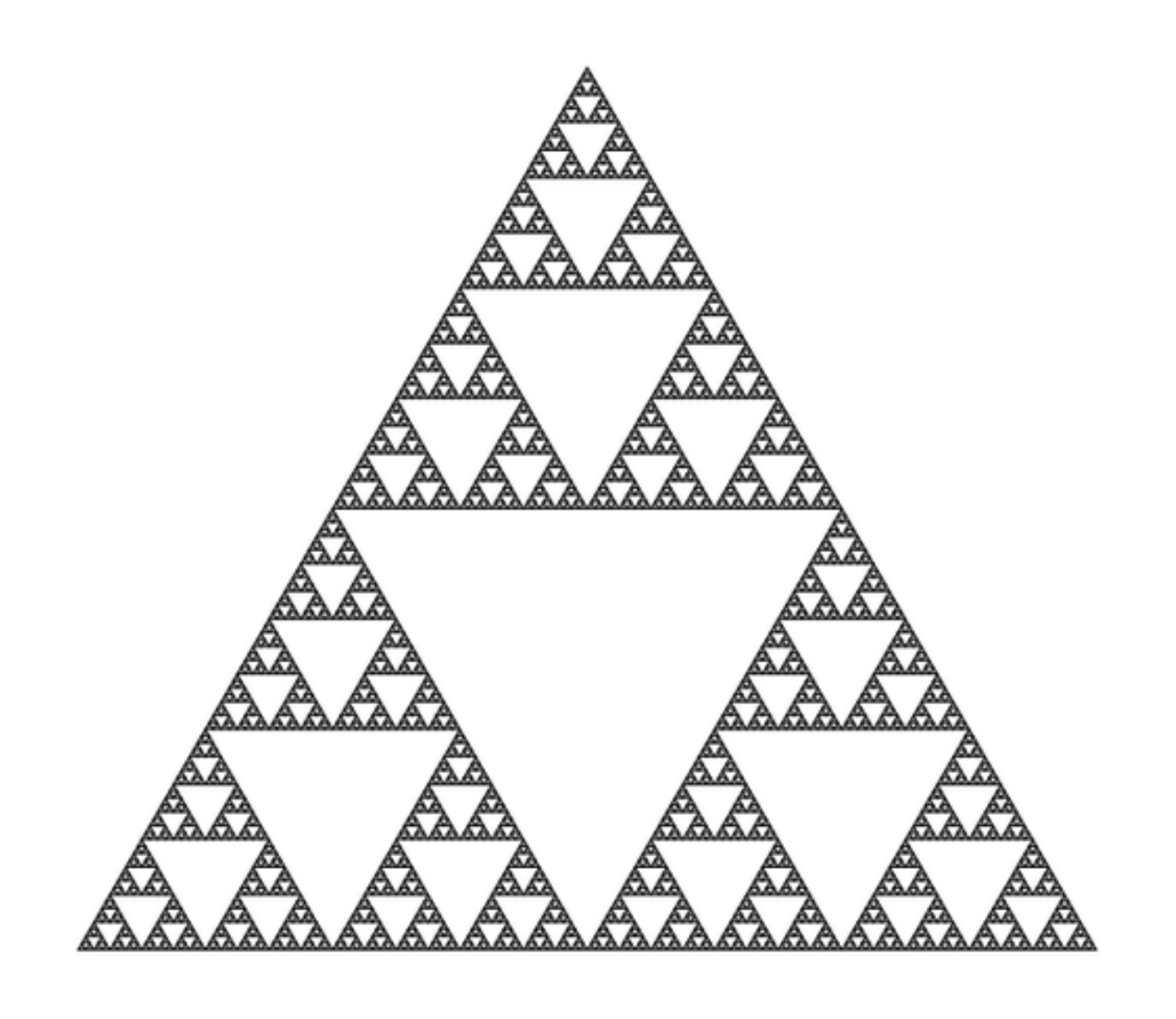}&\includegraphics[width=8cm]{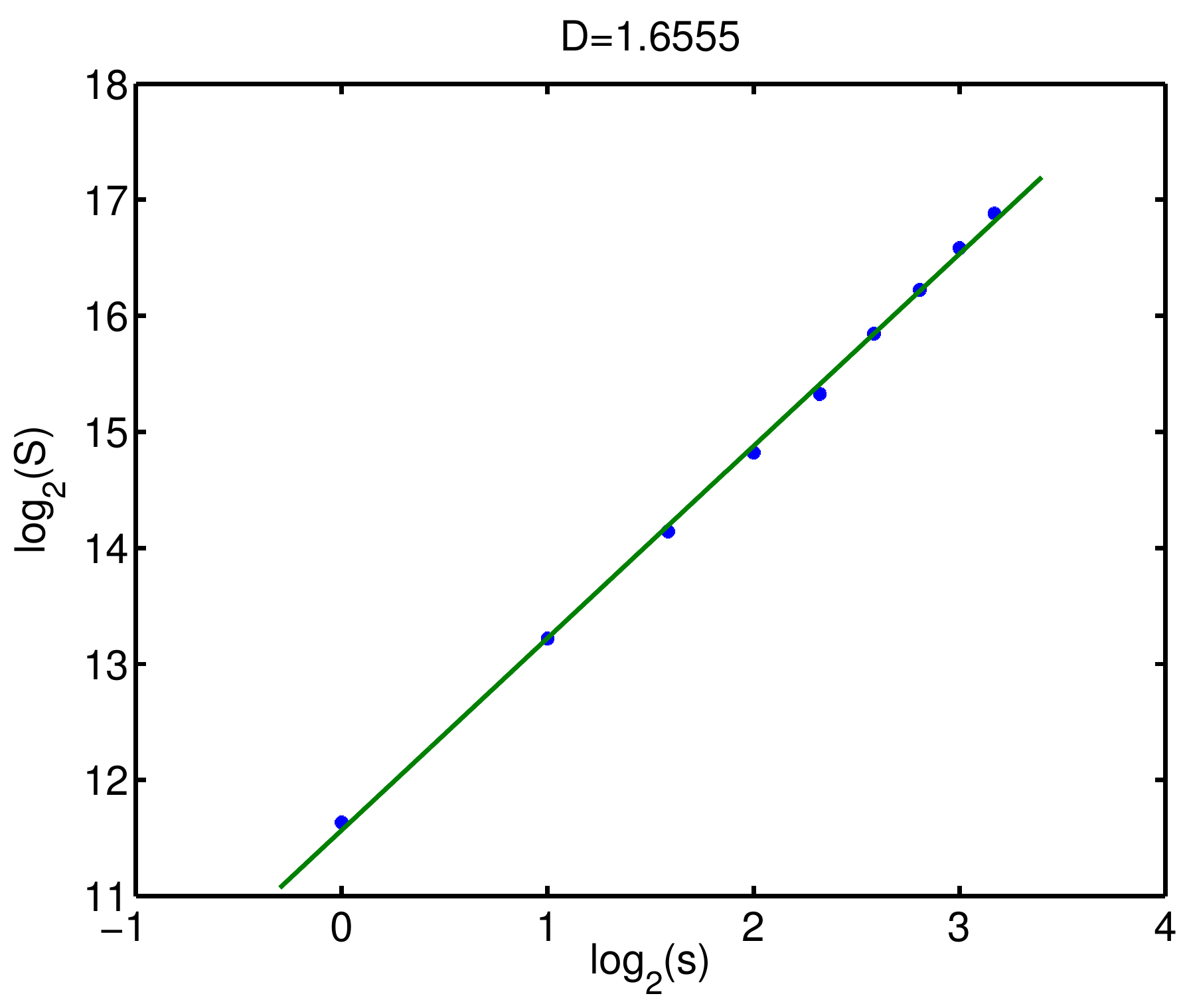}
\end{tabular}
\caption{(a) Sierpinski triangle and (b) its fractal dimension analysis.}\label{sierpinski}

\end{figure}

The fractals used for the analysis are displayed in figures \ref{cantor} to \ref{sierpinski}.  All the image files of the fractals that are analyzed have been downloaded from the Internet \cite{figswiki}.  The actual Hausdorff dimension listed in this web page has also been collected for comparison.  

In figures from \ref{cantor} to \ref{sierpinski}, each fractal to be analyzed is plotted at the left (a) panel and the result of the fractal dimension calculation is displayed in the right (b) panel.  In Table \ref{table},  the name of the fractal and the name of the file downloaded are listed, together with the actual Hausdorff dimension of the set analyzed and the dimensions computed following the algorithm described in this work both for grayscale $D_g$ and monochrome $D_{bw}$ scaled replicas of the original image.

\begin{table}
\begin{tabular}{p{0.2\linewidth}cp{0.4\linewidth}crcrcr}
\hline
\hline
Fractal name && File name & & $D_H$ & & $D_{g}$ & & $D_{bw}$\\
\hline
\hline
Asymmetric Cantor set && {\tt AsymmCantor.png} && $0.6942$ && $0.8320$ && $0.8754$\\
\hline
Boundary of the Dragon curve && {\tt Boundary\_dragon\_curve.png} &&$1.5236$ && $1.5946$ && $1.5225$\\
\hline
Fibonacci word fractal $60^o$ && {\tt Fibo\_60deg\_F18.png} && $1.2083$ && $0.7985$ && $0.7985$ \\
\hline
Ikeda map attractor && {\tt Ikeda\_map\_a=1\_b=0.9\_k=0.4\_p=6.jpg} && $1.7$ && $1.6687$ && $0.8681$\\
\hline
Julia set && {\tt Juliadim2.png} && $2$ &&  $1.8105$ && $1.7353$ \\
\hline
Julia set $z^2-1$ && {\tt Julia\_z2-1.png} && $1.2683$ && $1.2123$ && $1.7495$\\
\hline
Boundary of the L\'evy C curve && {\tt LevyFractal.png} && $1.9340$ && $1.9353$ && $1.9353$ \\
\hline
Sierpinski triangle && {\tt Sierpinski8.svg} && $1.5849$ && $1.6555$ && $1.3032$\\

\hline\hline
\end{tabular}
\caption{The five columns of the table correspond (from left to right) to the name of the fractal, the name of the file used\cite{figswiki}, the Hausdorff fractal dimension\cite{figswiki}, the computed fractal dimension obtained using black and white images at all scales and the computed fractal dimension obtained using grayscale images.}\label{table}
\end{table}

For almost all cases, the dimension calculated using the grayscale scaled images $D_g$ provides either equal or better accuracy than that given by the dimension calculated using the black and white scaled images $D_{bw}$.  It is noteworthy how our algorithm provides in most cases good approximation to  the exact dimension of the ideal object working with a, necessarily imprecise, representation of the mathematical object in an image file.  

The worst result is obtained for the Fibonacci fractal displayed in figure \ref{fibo}.  A detailed analysis shows that the image used in this case provides a rather poor representation for this fractal and the files corresponding to values of $s$ from $1$ to $3$ are actually blank.  This result could have been observed directly from the fractal dimension analysis shown in Fig. \ref{fibo} (b), where no change in the file size $S$ is obtained for these values of $s$.  Once these meaningless data points are eliminated from the analysis, the accuracy estimating the dimension improves, but is still far from the actual value.  A poor representation of the fractal complexity in the original image file can be inferred from this result.  

Another interesting example is provided by the Julia set $z^2-1$ displayed in figure \ref{juliaz} (a).  The analysis shows that the $s=1$ scaled version of the original image still has some information content, but the scaling analysis of figure \ref{juliaz} (b) displays a change in slope for the four data points corresponding to the lowest scales as compared with the tendency shown by the other points.  If these points are neglected, the estimate of the fractal dimension changes from $D=0.9590$ to $D=1.2123$, which is a significant improvement of the accuracy when compared with the actual value $D_H=1.2123$.  

\subsection{A systematic study based on the Weierstrass cosine function}

In this second analysis, a series of computer generated images has been used in order to perform a systematic evaluation of the proposed computational procedure.   The fractals have been generated using the Weierstrass cosine function \cite{esteller}
\begin{equation}
W_\alpha(t)=\sum_{n=0}^{M} \gamma^{-n\alpha}\cos\left(2\pi\gamma^nt\right),
\end{equation}
with $\gamma>1$ and $0<\alpha<1$.  For $M\to\infty$ the fractal dimension is $D=2-\alpha$ \cite{esteller}.  $\gamma$ and $M$ have been set to $\gamma=5$ and $M=26$, respectively, and values of $\alpha$ ranging from $0.2$ to $0.8$ have been used.

\begin{figure}
\begin{tabular}{cc}
(a)&(b)\\
\includegraphics[width=6cm]{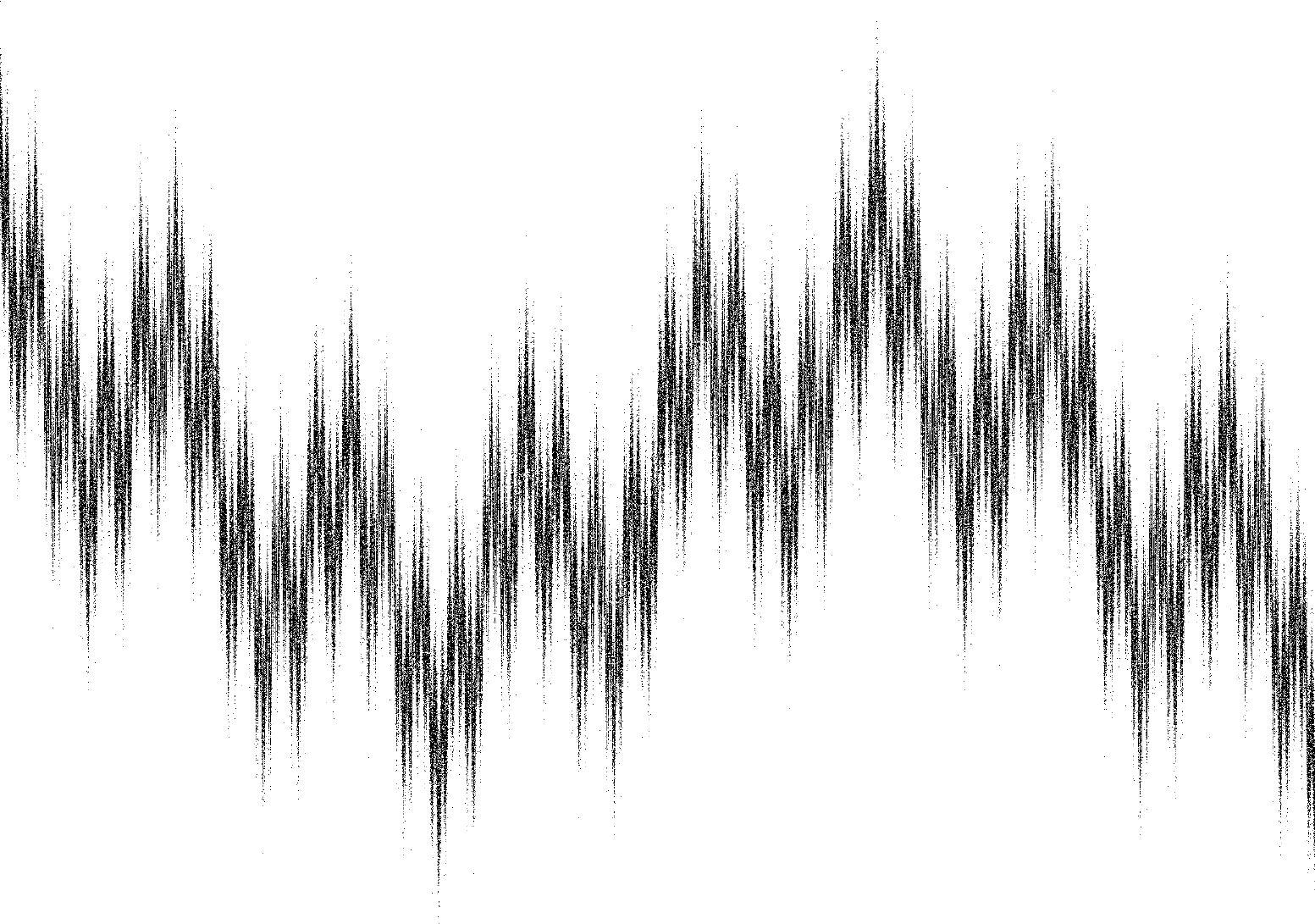}\hspace*{1cm}&\hspace*{1cm}\includegraphics[width=6cm]{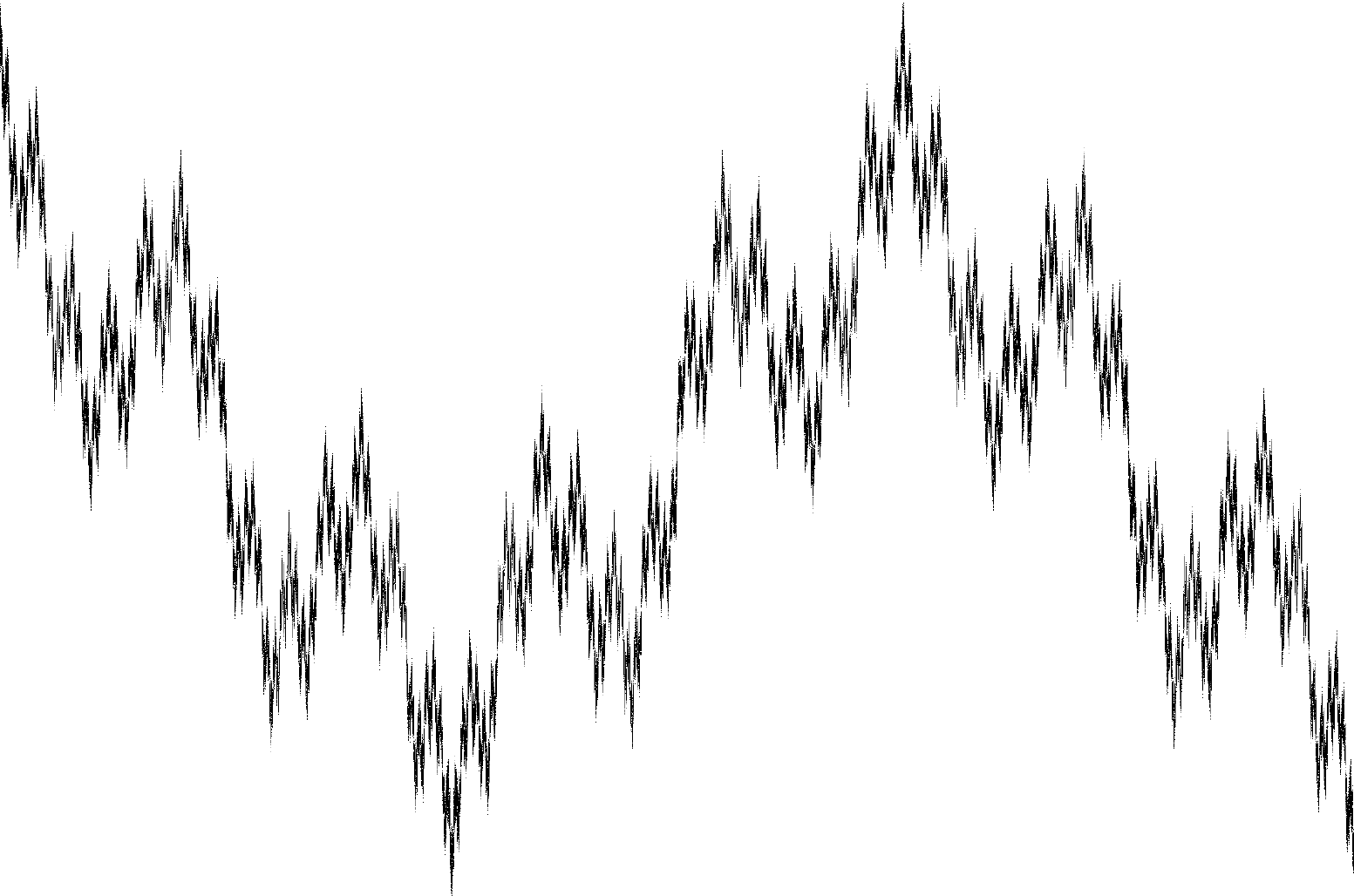}\\
(c)&(d)\\
\includegraphics[width=6cm]{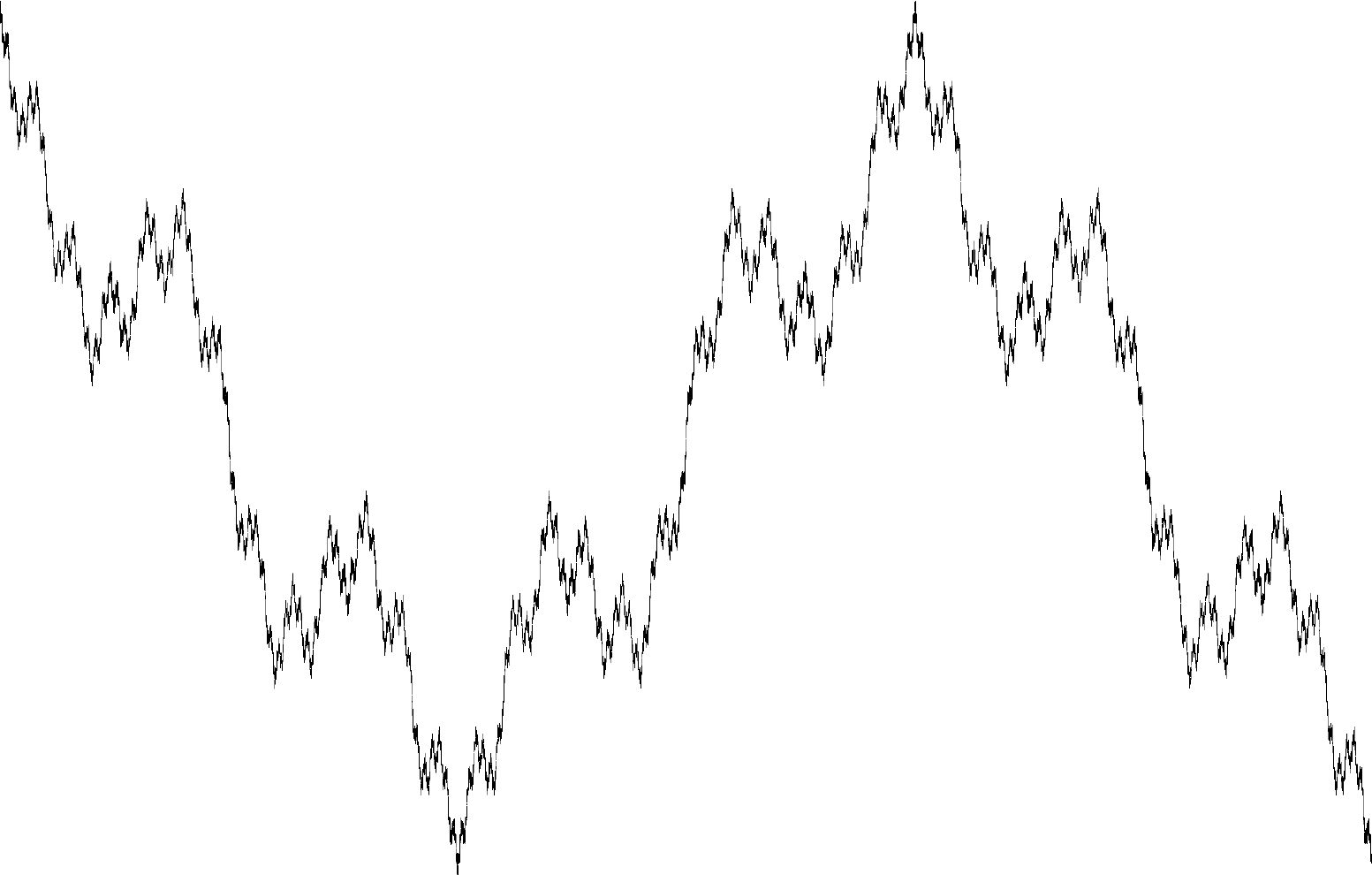}&\includegraphics[width=6cm]{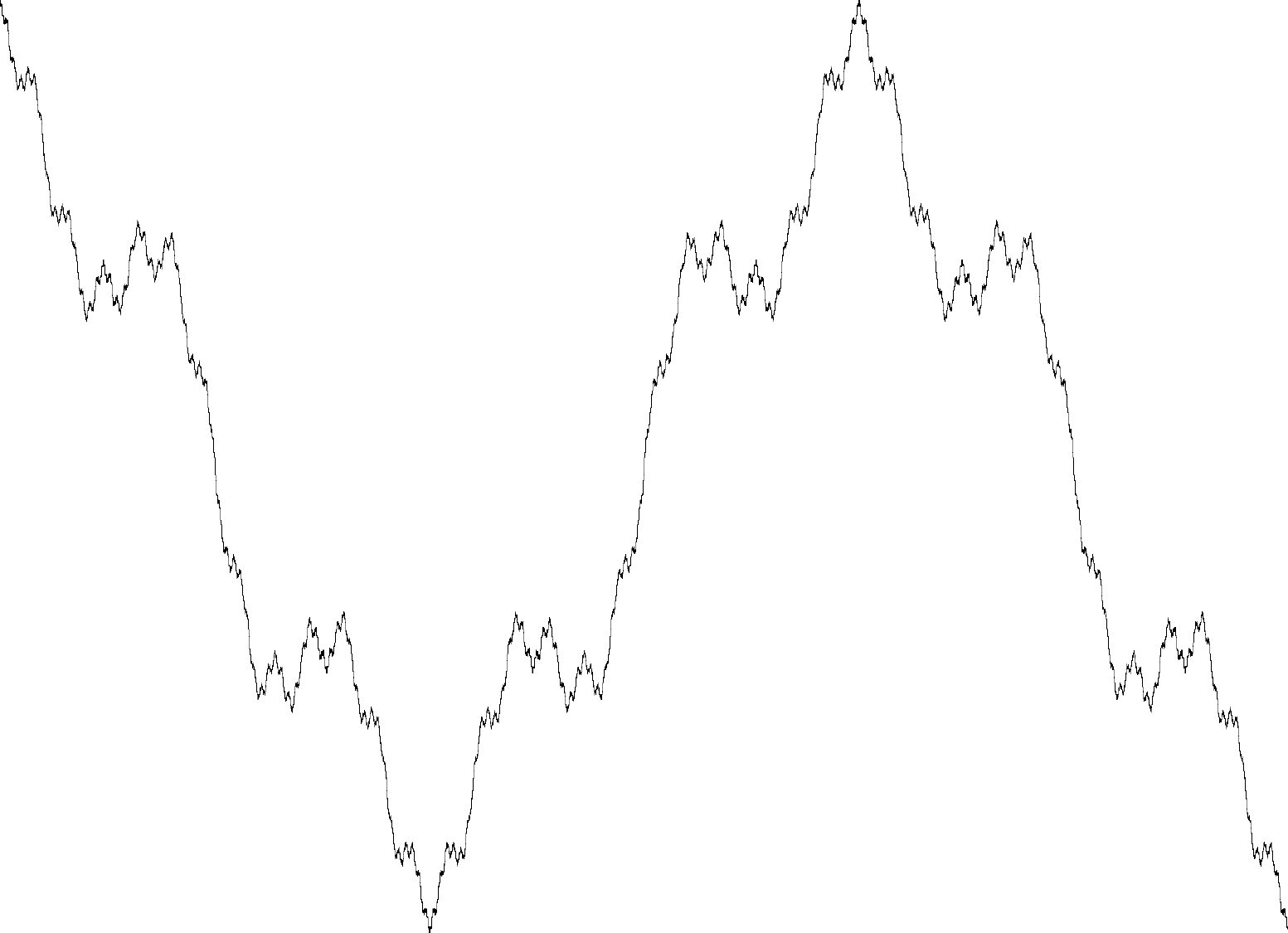}
\end{tabular}\caption{Four of the $W_\alpha$ fractal sets used in the study. (a) $\alpha=0.2$, (b) $\alpha=0.4$, (c) $\alpha=0.6$ and (d) $\alpha=0.8$.  The corresponding fractal dimensions are (a) $D=1.8$, (b) $D=1.6$, (c) $D=1.4$ and (d) $D=1.2$. }\label{imagenes}
\end{figure}

There follows the  details of each evaluation performed.  First,  for a given value of $\alpha$, $W_\alpha(t)$ is plotted with $N$ data points in the interval $t\in [0, 1.5]$ using {\tt Matlab}.  This plot is then printed to an uncompressed TIFF image file.   Four examples of the images used are displayed in Figure \ref{imagenes}. We stress that $N$ is the number of points in the {\tt Matlab} plot and not the number of pixels corresponding to the fractal in the image file.  The assignment of the values of the image pixels from the plotted data is internal to {\tt Matlab}.  The initial image generated with {\tt Matlab} contains $N_i=4800\times 36701=17284800$ pixels. The blank margins of the image are then removed using the {\tt ImageMagick} command {\tt mogrify -trim}.

A sequence of downscaled versions resized with percentages read from the vector 
\begin{equation}
V_s=[5\,\, 6\,\, 7\,\, 8\,\, 9\,\, 10\,\, 12\,\, 14\,\, 16\,\,  20\,\, 30\,\, 40\,\, 50\,\, 60\,\, 70\,\, 80\,\, 90]\label{vector}
\end{equation}
is generated from the initial image using ImageMagick's {\tt convert} program.  The values in \eqref{vector} have been arbitrarily chosen to produce a more or less regular spacing in the logarithmic plot.   As commented above, the scaling factor is set with the simplified version {\tt -scale} that reduces the processing in the downscaling to a pixel averaging \cite{imagick} instead of the {\tt -resize} option.  

The study has been repeated using sequences of downscaled images with PNG and TIFF image file formats.  In the case of TIFF images, two different types of compression have been used: LZW and ZIP.  The dimension calculations have also been repeated after externally compressing the sequences of image files using {\tt gzip} to check the existence of inefficiencies in the compression.  

The calculation of the compression dimension has been systematically repeated for different values of $\alpha$ and $N$.  In the analysis, it was frequently observed a  deviation from linearity at the larger values of $s$ in the fit of the $\log_2(S)$ vs $\log_2(s)$ data.   In order to quantify this effect, the fractal dimension has been calculated for different values of $n_s$, which is the number of elements of $V_s$ used, always starting from the smallest scale.  For instance, $n_s=8$ means that only the first eight values of $s$ in $V_s$ ($s=5,\, 6,\, 7,\, 8,\, 9,\, 10,\, 12,\, 14$) and the corresponding $S(s)$ data are used in the linear fit of the log-log plot for the calculation of the fractal dimension. 

\begin{figure}
\begin{tabular}{cc}
(a)&(b)\\
\includegraphics[width=8cm]{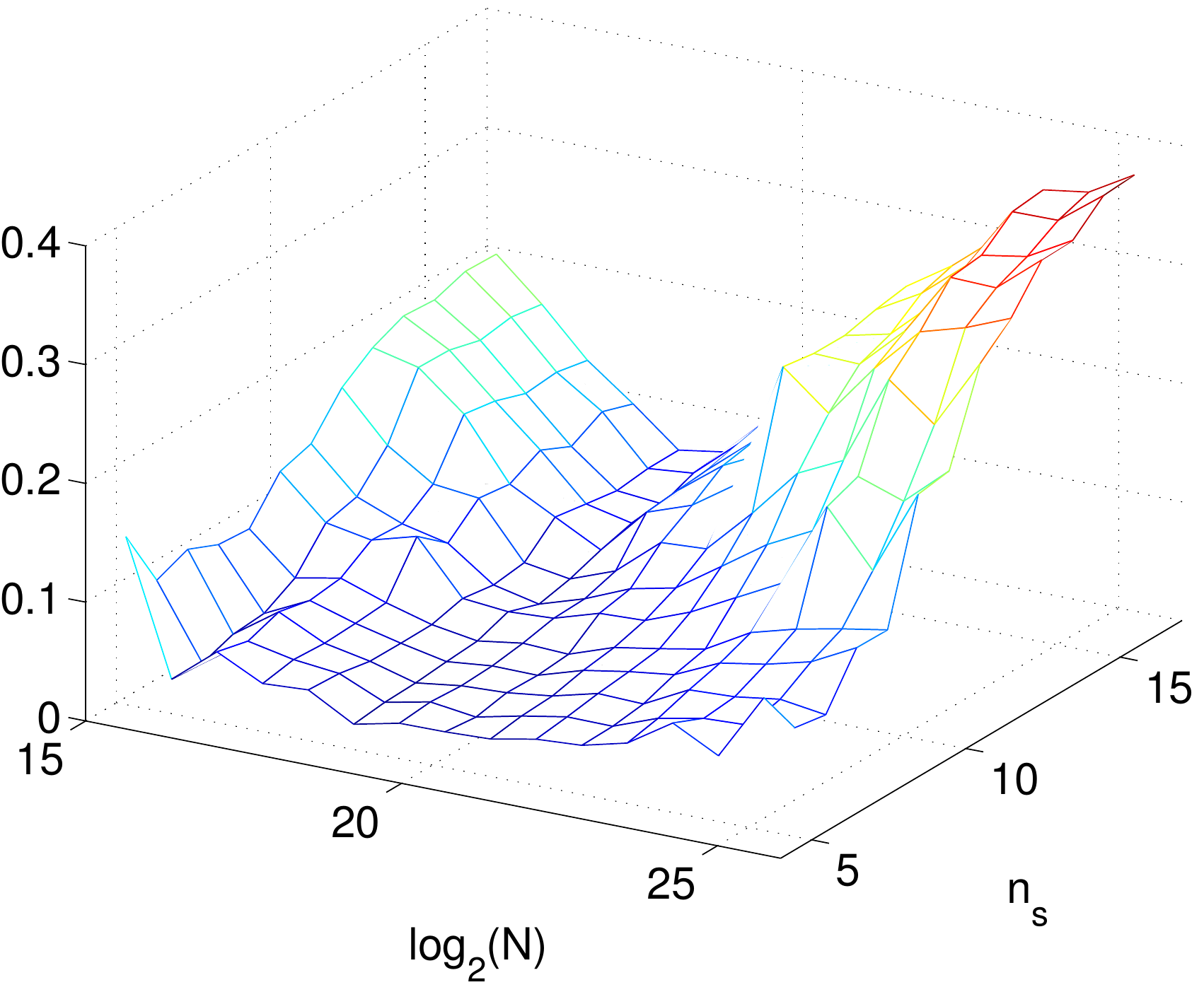}&\includegraphics[width=8cm]{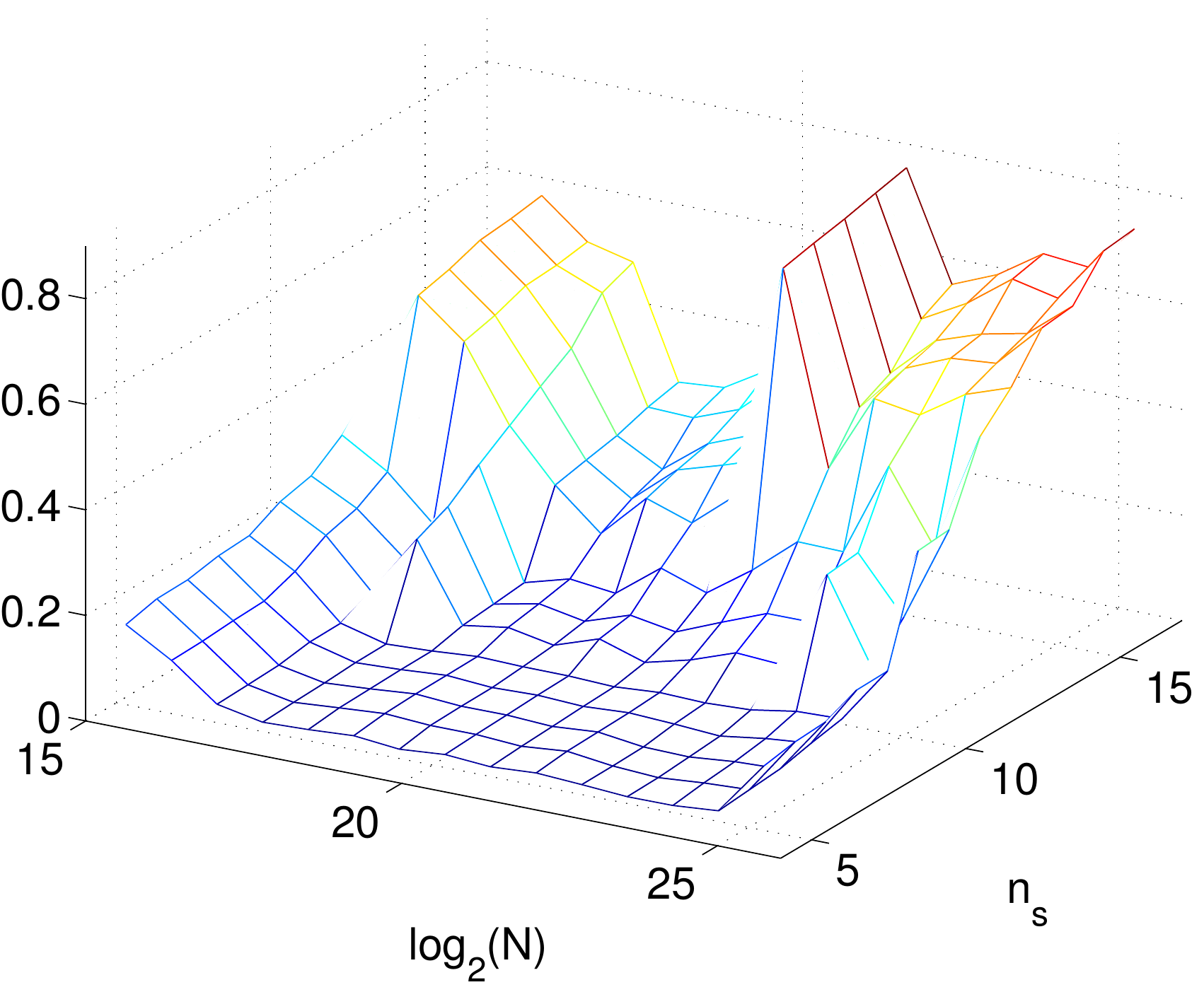}\\
\end{tabular}
\caption{(a) Unsigned mean error of the seven values of the fractal dimension (as $\alpha$ is varied from $0.2$ to $1.8$) calculated for each value of $N$ and $n_s$.  (b) Mean value of the norm of the residuals in the linear fit for the seven values of $\alpha$ at each $N$ and $n_s$.  PNG images have been used in the calculations.}\label{umePNG}
\end{figure}

\begin{figure}
\begin{tabular}{cc}
(a)&(b)\\
\includegraphics[width=8cm]{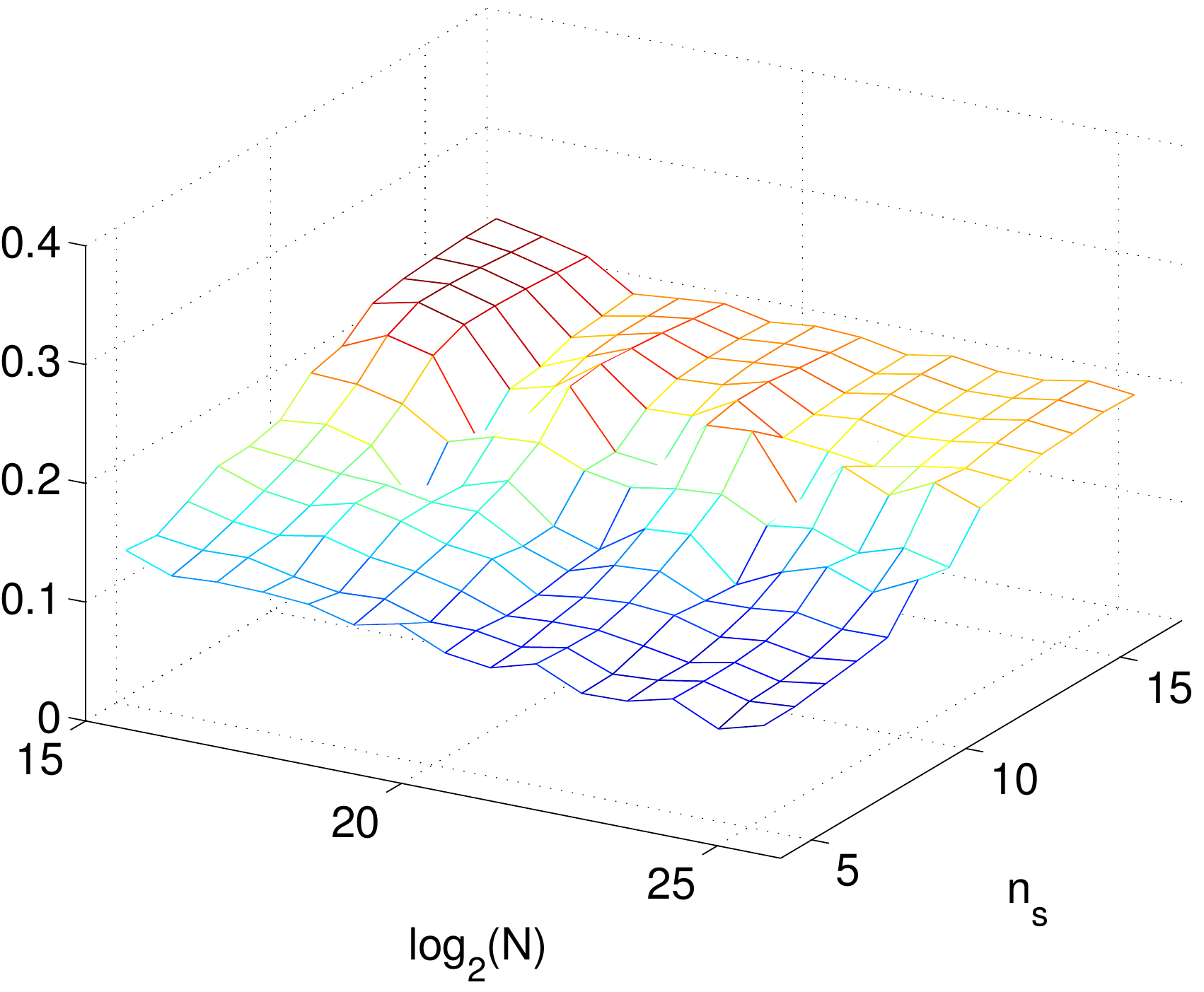}&\includegraphics[width=8cm]{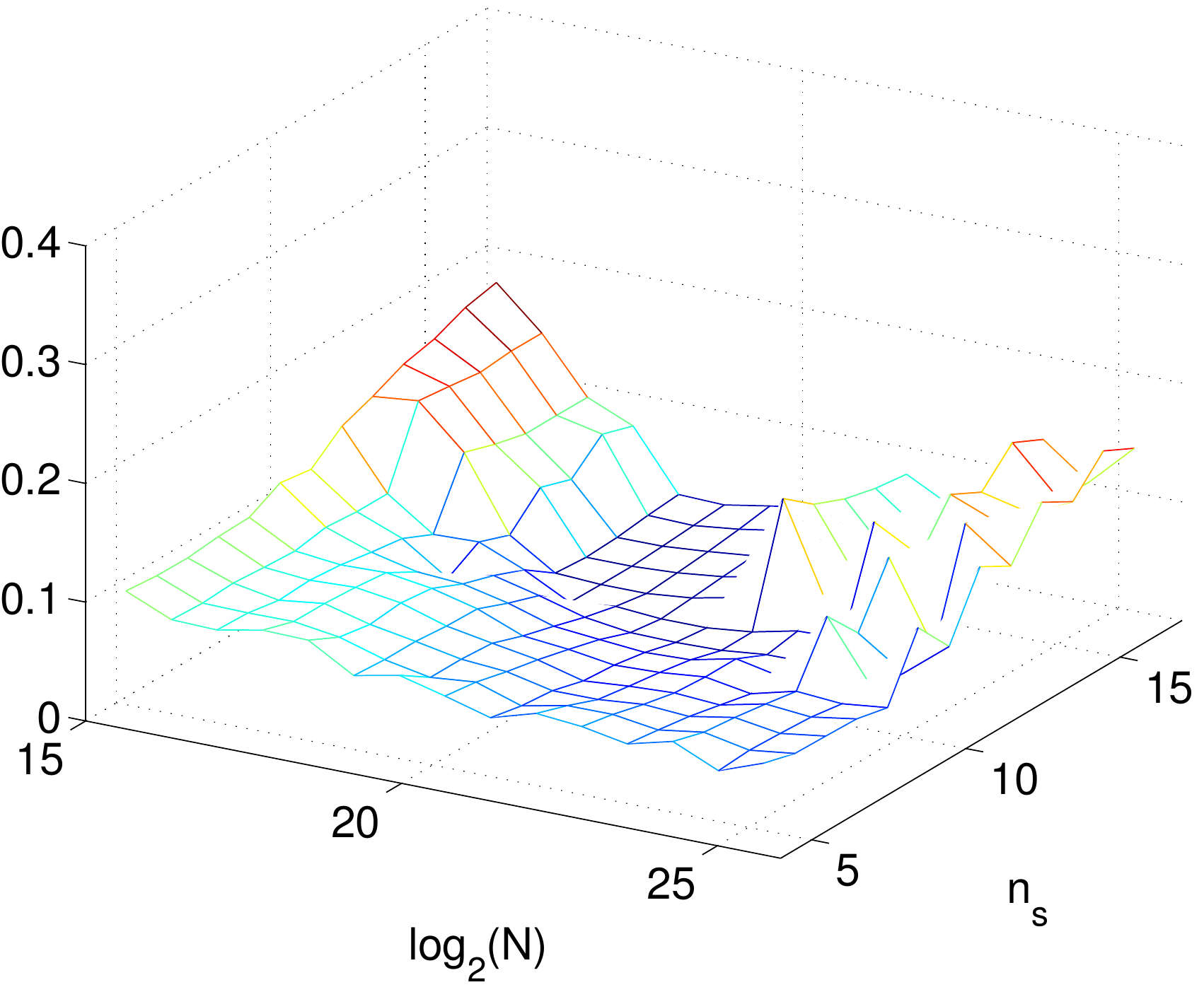}\\
(c)&(d)\\
\includegraphics[width=8cm]{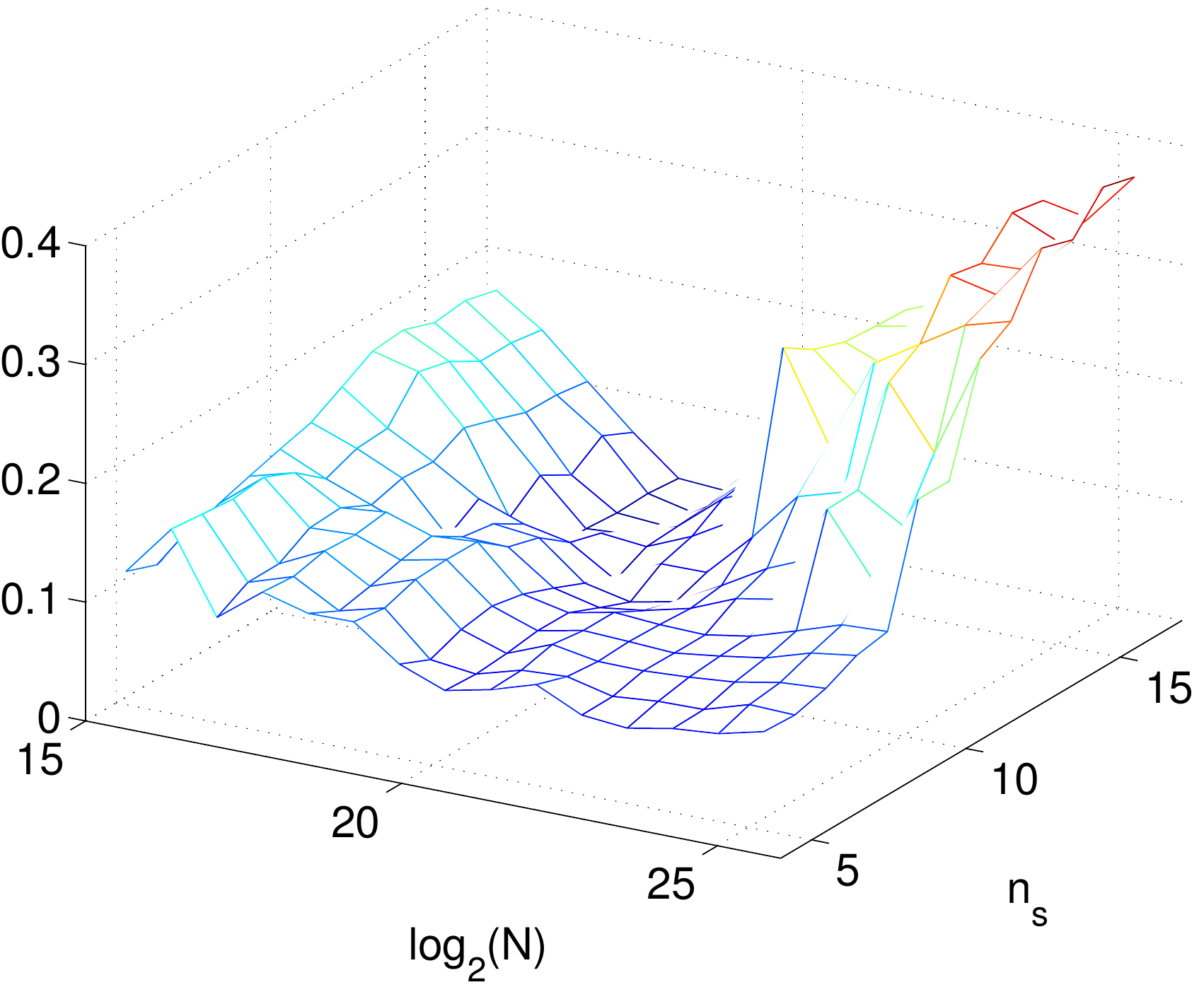}&\includegraphics[width=8cm]{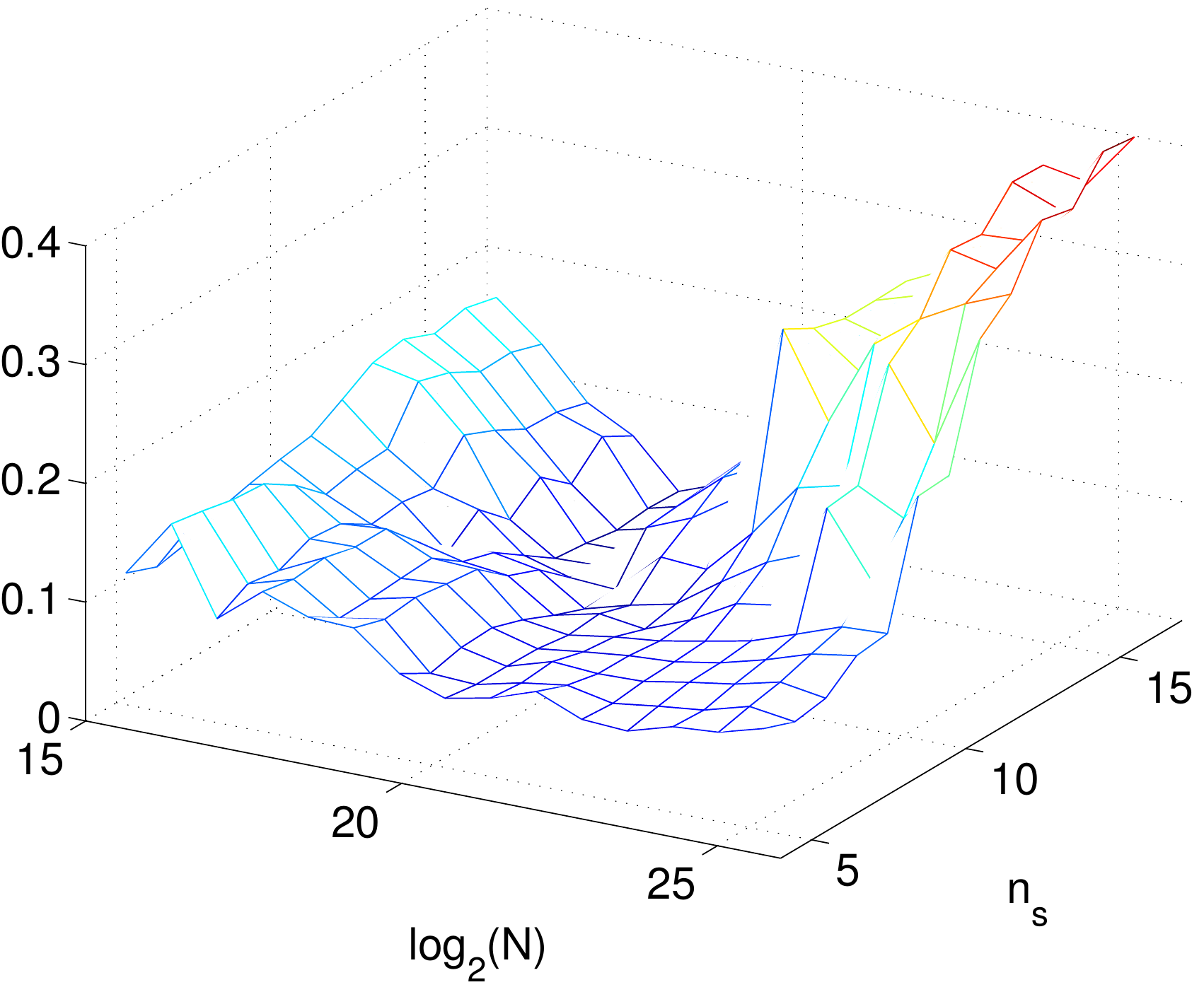}
\end{tabular}
\caption{(a) Unsigned mean error in the values of the fractal dimension computed using TIFF image format with LZW compression. (b) UME after a second compression using {\tt gzip}. (c) Unsigned mean error in the fractal dimension computed using TIFF image format with ZIP compression. (d) UME after a second compression using {\tt gzip}. }\label{umeTIF}
\end{figure}

\begin{figure}
\begin{tabular}{cc}
(a)&(b)\\
\includegraphics[width=8cm]{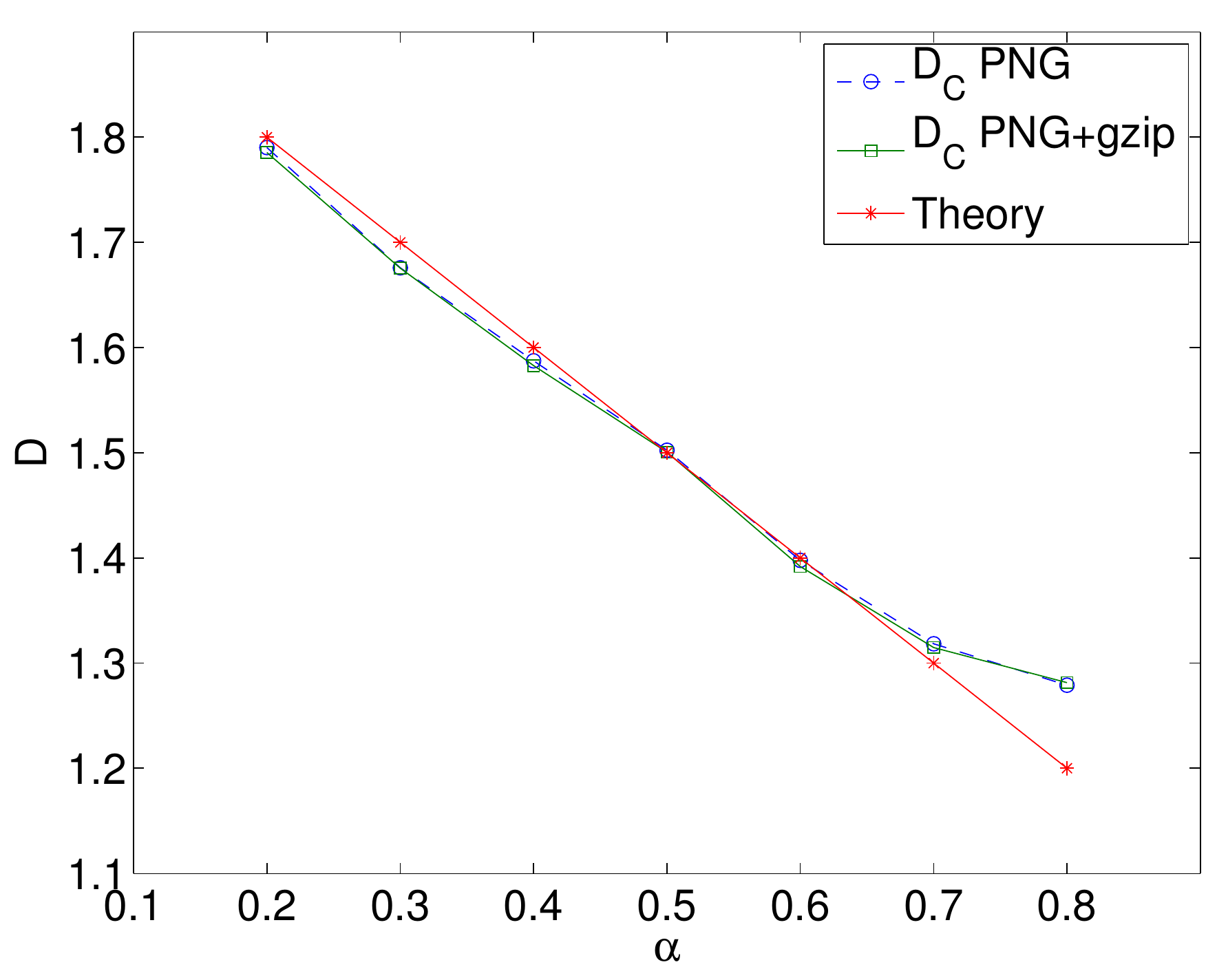}&\includegraphics[width=8cm]{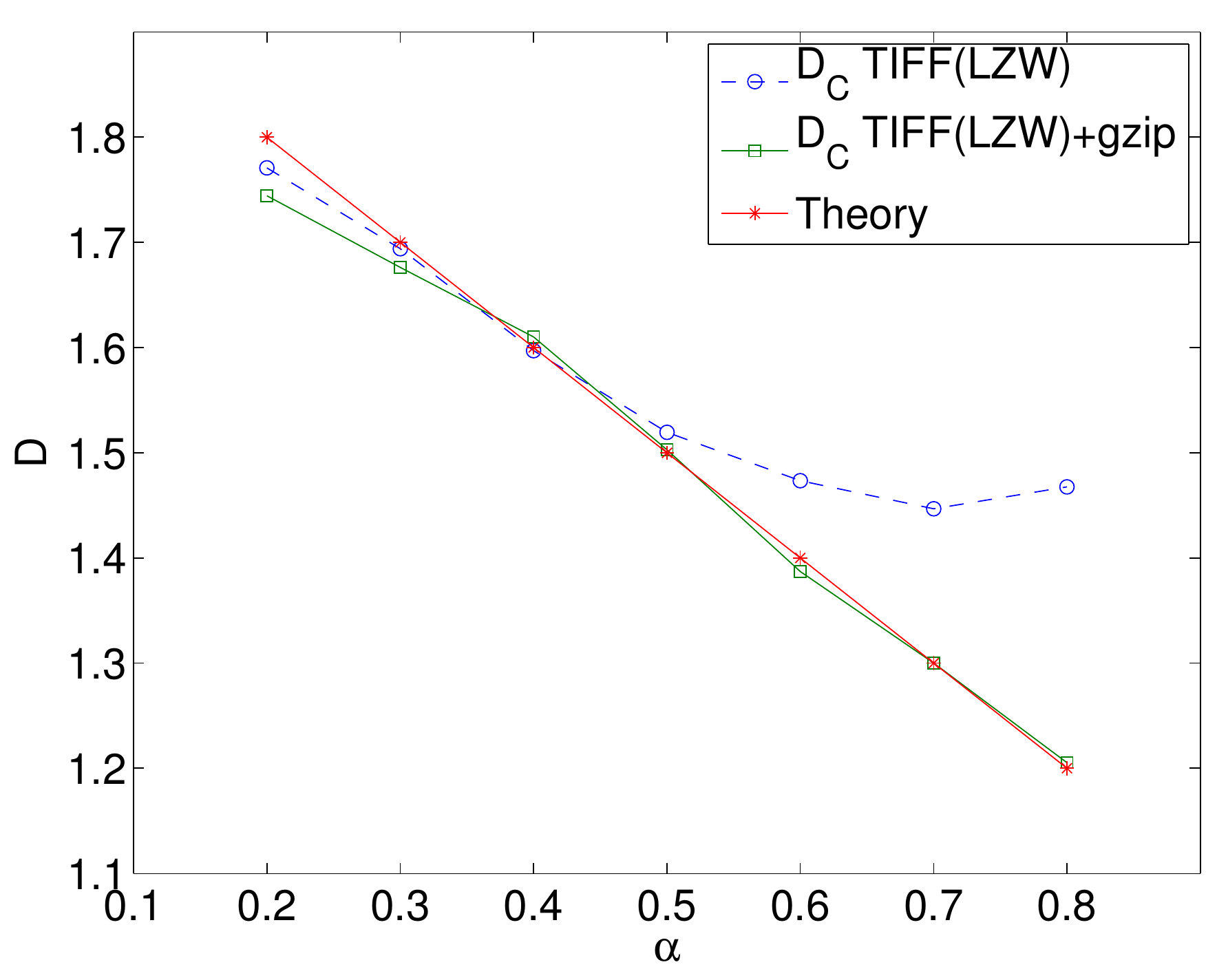}\\
(c)&(d)\\
\includegraphics[width=8cm]{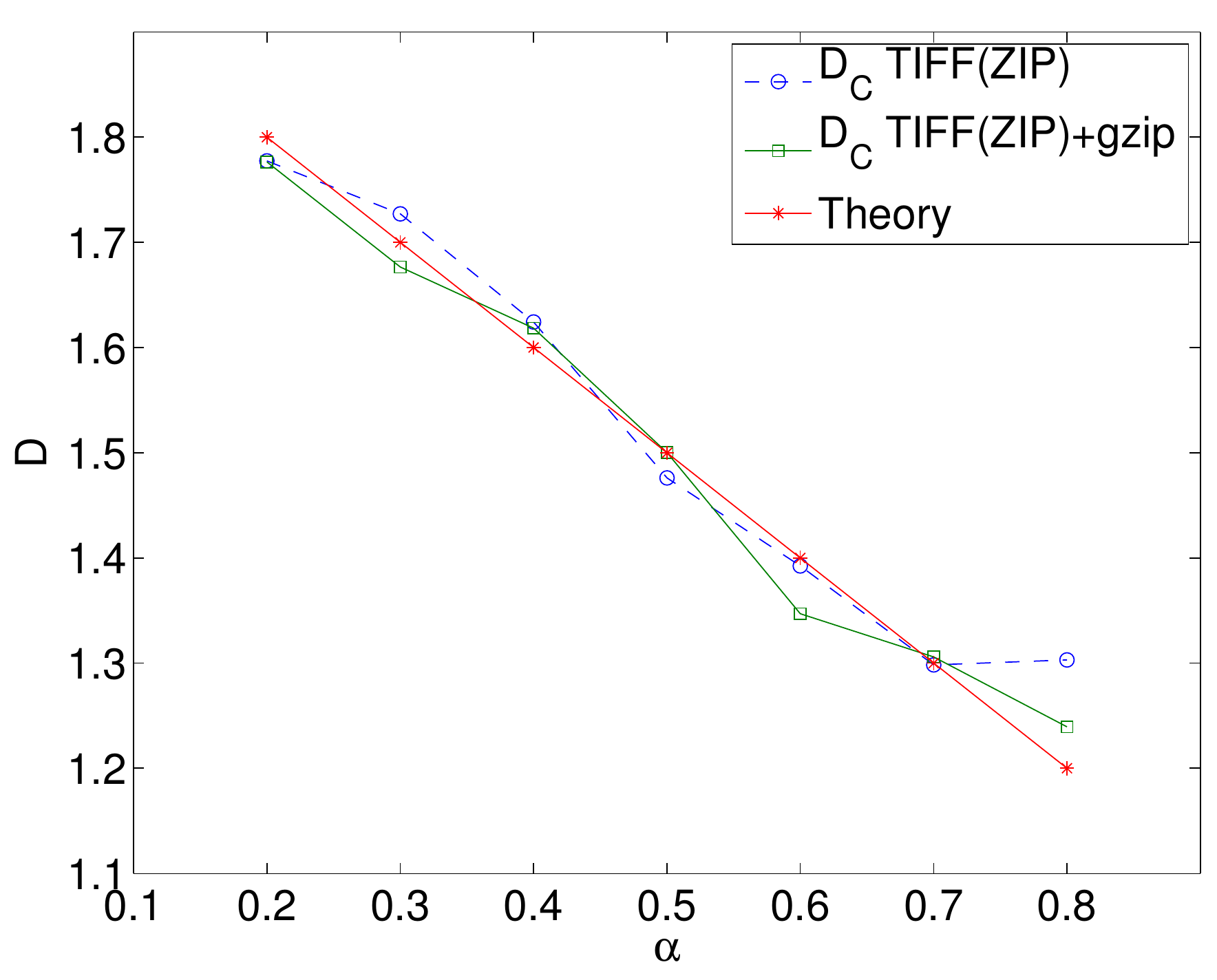}&\includegraphics[width=8cm]{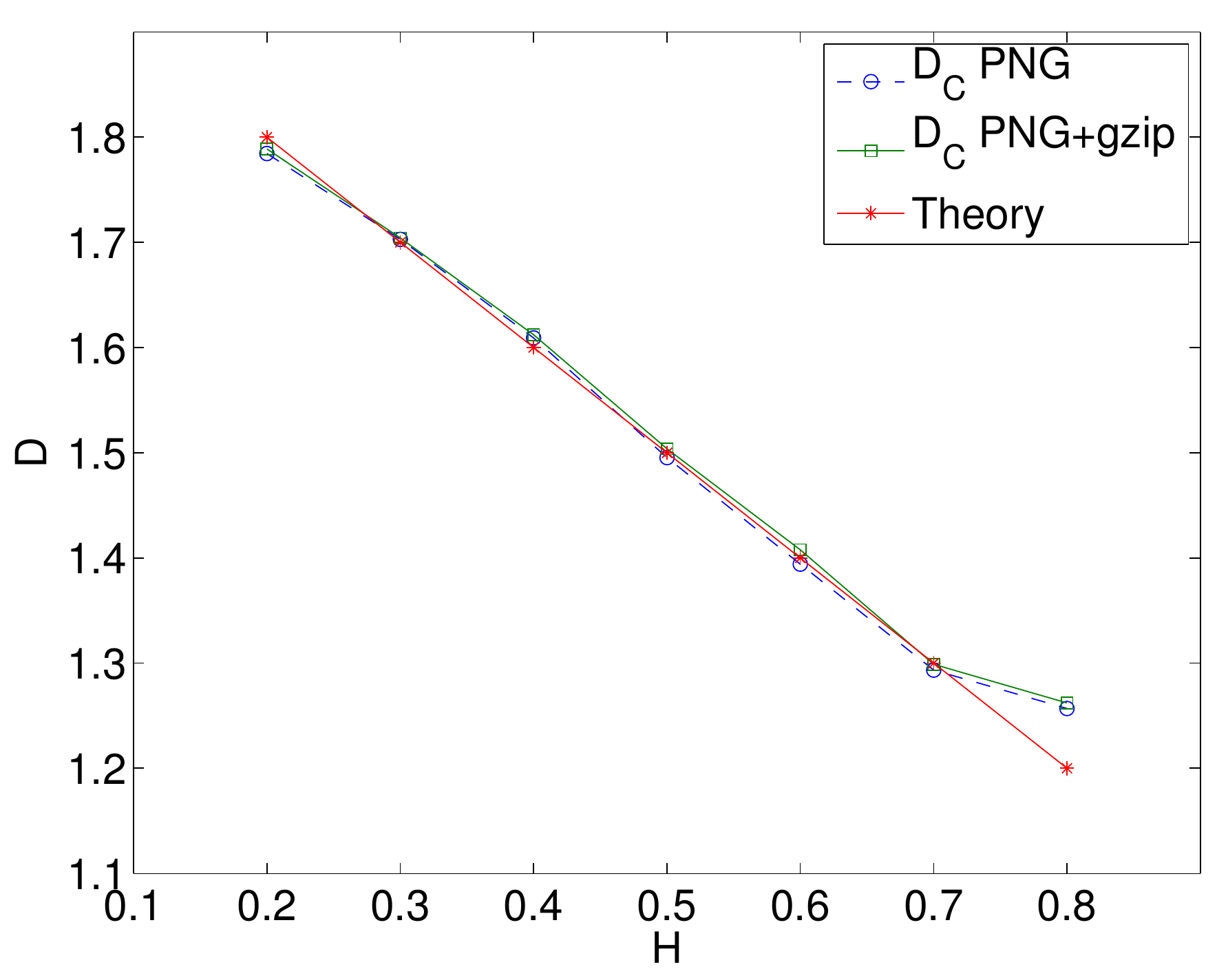}
\end{tabular}
\caption{Estimated dimensions at the minimum UME using (a) PNG format, (b) TIFF format with LZW compression, (c) TIFF format with ZIP compression and (d) PNG format with the \emph{-resize} {\tt ImageMagick} option instead of \emph{-scale}.  Asterisks correspond to the theoretical value of the fractal dimension, circles to the estimated dimension and squares to the dimensions estimated after a second compression using {\tt gzip}.  }\label{DC}
\end{figure}

For each value of $N$ and $n_s$, the compression fractal dimension has been calculated for seven values of $\alpha$ in the range between $0.2$ and $0.8$.  The unsigned mean error (UME) of these seven estimations is defined as
\begin{equation}
\text{UME}=\dfrac{1}{7}\sum_{l=1}^7\left|D_c(l)-D(l)\right|,
\end{equation}
where $\alpha(l)=0.2+(l-1)0.1$, $D_c(l)$ is the calculated compression dimension and $D(l)=2-\alpha(l)$ is the theoretical dimension.

The results obtained using PNG format images are plotted in Figure \ref{umePNG} (a).   The estimation error displays a characteristic dependence with the number data points in the {\tt Matlab } plot, with optimal performance at a given $N$.   These results exhibit the expected relation between the image resolution given by the number of image pixels $N_i$, the amount of information required to represent the fractal a given resolution level  $N_f=2^H$, and the number of data points initially used to plot the fractal $N$. For small $N$, the error of the estimation decreases as we increase $N$.  If we assume a direct relation between $N$ and the number of pixels corresponding to the fractal for the lowest values of $N$, an optimum should be obtained as $N \sim N_f$. At the same time, we need $N_f<<N_i$ for a faithful representation of the fractal and using an excessive number of data points saturates the image without adding more detail.  Therefore, the UME grows with $N$ after the optimum value is exceeded. Other tests performed with a reduced resolution of the initial image $N_i$ show the same qualitative behavior but with a corresponding reduction in the optimum value of $N$.  It is also noteworthy that the sensitivity of the error to the value of $N$ decreases when smaller values of $n_s$ are used and the data for the highest scales are neglected in the calculations.  This can be attributed to the incorporation of the fractal information in the grayscale coding along with the pixel averaging in successive downscaling stages, as discussed in Section III.

 Figure \ref{umePNG} (b) shows the average (over each set of seven values of $\alpha$) of the norm of the residuals in the linear fit used to determine the values of the fractal dimension. Large errors in the calculation of the fractal dimension in Figure \ref{umePNG} (a) are correlated with poor linear fits in the $\log_2(S)$ vs $\log_2(s)$ in Figure  \ref{umePNG} (b).  This reinforces the  consistency of the proposed method, since good fits (with small norm of residuals) to poor estimates seem not to be expected. 

The values of UME in the dimension calculation using TIFF images are shown in Figure \ref{umeTIF}.  Figure \ref{umeTIF} (a) shows the results obtained using LZW compression.  In general, the UME is substantial, specially for large values of $n_s$.  A second compression of the image files using {\tt gzip} produces significant size reduction factors when the image files are large.  This must be linked to a low efficiency of the LZW compressor for large file sizes \cite{Salomon}.  It can be corrected for with an external compression using {\tt gzip} prior to the computation of the fractal dimension, as shown in Figure \ref{umeTIF}(b).  Figure \ref{umeTIF}(c) shows the values of UME obtained using TIFF image format with ZIP compression.  These are comparable to those obtained with the PNG format and TIFF with LZW compression plus a second compression using {\tt gzip}.  Also, an external {\tt gzip} pass to the TIFF files with ZIP compression does not produce an improvement of the estimation error in Figure \ref{umeTIF} (d).

For each file format, the dimensions calculated with the best UME are shown in Figure \ref{DC}, together with the theoretical values and the estimations obtained after a second external compression using {\tt gzip}.  The results using PNG images are very good except for the smallest dimension considered.  A second compression shows no effect on the results.  When TIFF images with LZW compression are used, very large errors are obtained when $D<1.5$, but these are due to the aforementioned poor performance of the LZW compressor when the image files are large and are corrected for using a second compression.  The accuracy is then excellent except for $D=1.8$.  Using TIFF images with ZIP compression yields reasonable estimates except for $D=1.2$.  The details of the downscaling of the image file also affect the estimation of the fractal dimension.  This is illustrated in Figure \ref{DC} (d) where PNG images are used but the {\tt ImageMagick} {\tt -resize} option is used instead of {\tt -scale}.  In this case, nearly exact values of the fractal dimension are obtained except for $D=1.2$.  

It is interesting to note that worst estimation results tend to show up either at the extreme values of $D$, closest to $D=2$  (the space filling plot) and $D=1$ (the case of an ordinary curve in Euclidean space) where the fractal complexity is probably most difficult to capture in an image file.   

A comparison between the performance of different algorithms commonly used to calculate the dimension of fractal waveforms tested with the Weierstrass cosine function was presented in \cite{esteller}.  Even though the methods studied in \cite{esteller} are applied to fractal data sequences and, therefore, are completely different to the computational scheme of this work, which acts of image files, it is interesting to note that the results from the method presented here are comparable, in terms of accuracy, to the most accurate of the methods analyzed in \cite{esteller}, namely, Highuchi method.

\section{Conclusion}

A method to calculate the information dimension of a fractal based on data compression has been presented.  An experiment has been set-up using images of fractal sets downloaded from the Internet and freely available software.  The results show good agreement in the estimated dimension and the exact values when the image file reproduces enough detail of the geometrical object under study.  The proposed scheme is particularly simple and it is even suitable for a hands-on introductory approach to concepts in information theory, fractal geometry and complexity.  In a more extensive analysis of the algorithm applied to the Weierstrass cosine function, the accuracy of the calculated dimension is found to be comparable to that of the methods normally employed in the estimation of the dimension of fractal sequences.

\section*{Acknowledgments}

This work has been funded by MINECO/FEDER grant no. TEC2015-69665-R and JCyL grant no. VA089U16.

\end{document}